\documentclass[useAMS,usenatbib]{mn2e}
\usepackage{times}
\usepackage{graphicx, amssymb}
    
\newcommand{\km}{${\rm km\,s}^{-1}$}
\newcommand\nodata{ ~$\cdots$~ }%

\newcommand{\fuse}{{\em FUSE}}

\newcommand{\hi}{H$\;${\small\rm I}\relax}
\newcommand{\hii}{H$\;${\small\rm II}\relax}
\newcommand{\heii}{He$\;${\small\rm II}\relax}

\newcommand{\ari}{Ar$\;${\small\rm I}\relax}

\newcommand{\cli}{Cl$\;${\small\rm I}\relax}
\newcommand{\cii}{C$\;${\small\rm II}\relax}

\newcommand{\civ}{C$\;${\small\rm IV}\relax}

\newcommand{\niii}{Ni$\;${\small\rm II}\relax}
\newcommand{\nv}{N$\;${\small\rm V}\relax}
\newcommand{\oi}{O$\;${\small\rm I}\relax}

\newcommand{\ovi}{O$\;${\small\rm VI}\relax}

\newcommand{\sii}{S$\;${\small\rm II}\relax}
\newcommand{\siii}{Si$\;${\small\rm II}\relax}
\newcommand{\siiii}{Si$\;${\small\rm III}\relax}
\newcommand{\siiv}{Si$\;${\small\rm IV}\relax}

\newcommand{\mgii}{Mg$\;${\small\rm II}\relax}

\newcommand{\feii}{Fe$\;${\small\rm II}\relax}
\newcommand{\feiii}{Fe$\;${\small\rm III}\relax}

\newcommand{\pv}{P$\;${\small\rm V}\relax}

\title[Highly ionised species in the LMC]{Highly ionised plasma in the Large Magellanic Cloud: 
Evidence for outflows and a possible galactic wind\thanks{Based on observations made with the NASA-CNES-CSA 
Far Ultraviolet Spectroscopic Explorer. FUSE is operated for NASA by the Johns 
Hopkins University under NASA contract NAS5-32985. Based on observations made with the NASA/ESA Hubble Space Telescope,
obtained at the Space Telescope Science Institute, which is operated by the
Association of Universities for Research in Astronomy, Inc. under NASA
contract No. NAS5-26555.}
}
\author[N.~Lehner \& J.C.~Howk]{N.~Lehner \ 
 \& J.C.~Howk
\\
Department of Physics, University of Notre Dame, 
225 Nieuwland Science Hall, Notre Dame, IN 46556, USA}
\date{Accepted XXX.
      Received XXX.}

\pagerange{\pageref{firstpage}--\pageref{lastpage}}
\pubyear{2007}

\begin{document}

\maketitle

\label{firstpage}

\begin{abstract}
Based on an analysis of the interstellar highly ionised species \civ, \siiv, 
\nv, and \ovi\ observed in the {\em FUSE}\ and {\em HST}/STIS E140M spectra of four hot stars in the 
Large Magellanic Cloud (LMC), we find evidence for a hot LMC halo fed by energetic outflows from the LMC 
disk and even possibly an LMC galactic wind. Signatures for such outflows 
are the intermediate and high-velocity components 
($v_{\rm LSR} \ga 100$ \km) relative to the LMC disk observed in the high- and low-ion 
absorption profiles. The stellar environments produce strong, narrow ($T\la 2\times 10^4$ K) 
components of \civ\ and \siiv\ associated with the LMC disk; in particular they are likely signatures
of \hii\ regions and expanding shells. Broad components are observed in 
the profiles of  \civ, \siiv, and \ovi\ with their widths implying  hot, collisionally ionised gas 
at temperatures of a few times $10^5$ K. There is a striking similarity in the \ovi/\civ\ 
ratios for the broad LMC and high-velocity components, suggesting much of the material at $v_{\rm LSR} \ga 100$ 
\km\ is associated with the LMC. The velocity of the  high-velocity component is 
large enough to escape altogether the LMC, polluting the intergalactic space between the LMC and the Milky Way. 
The observed high-ion ratios of the broad LMC and high-velocity  components are consistent
with those produced in conductive interfaces; 
such models are also favored by the apparent kinematically coupling between the high and the weakly ionised species.
\end{abstract}


\section{Introduction}
The structure and energetics of the interstellar medium (ISM) in
galaxies are strongly influenced by stellar winds and supernovae
(e.g., Abbott 1982).  This is particularly true for spiral and
irregular galaxies with on-going massive star formation, such as the
Milky Way, and the Large and Small Magellanic Clouds (LMC, SMC).  The
input of energy from winds and supernovae is expected to affect the
ISM on kiloparsec scales, creating regions of super-heated gas (e.g.,
Norman \& Ikeuchi 1989). The $10^6$ K gas
that expands into the halo of a disk galaxy may eventually cool and
fall back onto the disk in a massive system, participating in a
galactic fountain (Shapiro \& Field 1976); in a low mass system,
this material may escape the galaxy altogether in a galactic wind
\citep{breit91}. Galactic fountains and winds affect the scale over which newly
produced metals are distributed in the ISM of a galaxy and/or in the
intergalactic medium. For example, outflows can affect the [$\alpha$/Fe] ratios 
inside and outside dwarf galaxies \citep{recchi01}. Gas dynamics are therefore 
critical in determining the evolution of galaxies \citep[e.g.,][]{matteucci03}. 
Characterising the phenomena of infall and outflow in nearby galaxies is 
crucial for understanding the role they may play in galaxy evolution and in the enrichment and 
evolution of the intergalactic medium.  

The LMC is the nearest gas-rich disk galaxy
to our own Milky Way.  As such it has been the target of intense
studies of its interstellar medium (ISM), including its dust (e.g.,
Meixner et al. 2006), its molecular gas (Tumlinson et al. 2002), its
warm \hi\ (Kim et al. 2003), its warm ionised gas traced by
H$\alpha$ emission (e.g., Danforth et al. 2002), its warm-hot gas
traced by \ovi\ (Howk et al. 2002; Danforth \& Blair 2006), and its
hot gas traced by X-rays (e.g., Snowden \& Petre 1994). In this 
work, we report observations of interstellar absorption from the Li-like ions
\nv, \civ, \siiv, and \ovi\ towards stars within the LMC. These
ions give us the means to study the highly-ionised material produced by 
feedback processes in the Milky Way and in the Magellanic Clouds. 
\siiv, \civ, \nv, and \ovi\ peak in abundance at 
(0.6,1.0,1.8, and 2.8)\,$\times$$10^5$ K, respectively, in collisional ionisation equilibrium 
(CIE; Sutherland\& Dopita 1993).  They are, however, unlikely to be in equilibrium
since these temperatures correspond to the peak of the interstellar
cooling curve, and gas at those temperatures is likely to cool faster
than it recombines. Instead, these ions may trace interfaces between very
hot gas ($T \ga 10^6$ K) and warm or cool ($T \la 10^4$ K)
material and serve as a probe of the interaction 
between the phases of the ISM having the most mass (warm/cool gas) and 
the most direct connection to feedback processes (hot gas).

Observational searches for a hot corona about the LMC using the highly ionised species
started with  {\em IUE}\ spectra of \civ, \siiv, and \nv\ towards the hot star 
Sk$-$67\degr104 \citep{deboer80}. 
However, the crude spectral resolution of {\em IUE}\ and the fact that
that the hot star could be partly responsible for producing \civ\ and \siiv\ 
brought \citet{chu94} to conclude these observations were not sufficient
to prove the existence of a hot LMC halo (\nv\ was only marginally detected
at best). These highly ionised species could be either
produced by photoionisation from the star itself or in the superbubble via
shocks. \citet{wakker98} found \civ\ absorption in the spectra of cooler 
stars (O9.7--B1) using modest resolution {\em HST}\ spectroscopy. 
This was the first evidence of \civ\ not related directly 
to local effects, as these stars were deemed unable to produce 
significant high-ion absorption via photoionisation.

More recently, {\em FUSE}\ has contributed to the effort of examining
the hot gas in the LMC by probing diffuse \ovi\ in both
Magellanic Clouds \citep{howk02,hoopes02,danforth02,danforth06}.  
\citet{howk02} used  {\em FUSE}\ observations of sight lines to the LMC to study the \ovi\
content and kinematics of the LMC.  Their survey of 12 sight lines 
through the near side of the LMC probed collisionally-ionised
gas at $T\sim 3\times 10^5$ K in the LMC and shows in particular that the LMC contains a
significant amount of highly-ionised gas,  with \ovi\ column densities
equal or greater than those found along extended paths through the
Milky Way; and much of the \ovi\ in the LMC must reside in an extended
``halo'' or thick disk distribution, likely similar to that found in
our Galaxy \citep{savage03}.

These studies of the individual highly ionised species have broadly sketched a picture
of the hot gas in the LMC and the presence of coronal hot gas, but
a crucial step is to combine all the available highly ionised species observed with 
{\em HST}/STIS E140M and {\em FUSE}\ for a better understanding of the highly ionised gas in 
the LMC. In this paper, we take such a step by presenting the detailed
kinematics and high-ion absorption (\ovi, \civ, \siiv, and
\nv) towards four stars in the LMC using {\em FUSE}\ and STIS.  Two of
these sight lines probe prominent superbubbles, while the other two
probe \hii\ regions, allowing us to compare the highly ionised species and their 
connection with weaker ions towards different types of interstellar regions, 
which is useful for  differentiating local from 
global effects.  The organisation of this paper is as follows. After describing 
the observations, data reduction, and analysis of the data (measurements
of the kinematics, column densities, and column density ratios)  in \S\ref{sec-obs}, we 
present an overview of the kinematics in \S\ref{sec-kin}.  In \S\ref{origint} we 
discuss the origin of the highly ionised plasma in the LMC and the implications of our 
results. A summary of the main results is presented in \S\ref{sec-sum}.

\section{Observations and analysis}\label{sec-obs}
\subsection{Database}
From the data archives of {\em FUSE}\ and {\em HST}/STIS E140M, we selected
four LMC targets. The selection criteria were to have the four critical 
highly ionised species (\siiv, \civ, \nv, and \ovi) observable in the FUV and UV  and
to be able to model these ions without too much uncertainties
(i.e. with a continuum near the relevant ions that can be modeled without too
much uncertainty). This resulted
in the sample of the four stars presented in Table~\ref{t1}. Fortuitously,
these stars are also located in different interstellar environments (from 
Chu et al. 1994 and references therein):
\\
(1) Sk$-$65\degr22 is projected within the supergiant shell LMC\,1,
which is about 1 kpc across \citep{meaburn80}. It is
located in a faint \hii\ region DEM\,48 at the southeast rim. 
This supergiant shell contains the OB association LH\,15.
\\
(2) Sk$-$67\degr211 is in the \hii\ region DEM\,241 that contains the OB association LH\,82.
The bright UV continuum and the early spectral type (O2\,III(f$^*$), Walborn et al. 2002) 
imply the star can likely produce strong \civ\  and \siiv\ absorption associated with its stellar
environment. 
\\
(3) Sk$-$69\degr246 is a WR star situated in the X-ray bright superbubble 30\,Dor, 
and more precisely is in the south end of Shell 5, which is the X-ray-bright "chimney"
to the north of the LH\,100 OB association (Y.-H. Chu 2006, priv. comm.).
\\
(4) Sk$-$71\degr45 is located in the superbubble DEM\,221 associated with the 
OB association LH\,69.

\begin{table*} 
\begin{minipage}{15.7 truecm}
\caption{\large Summary of stellar and sightline properties \label{t1}}
\begin{tabular}{lllcccccl}
\hline
Name    & Alias  &   Type  &   $l$&   $b$&$V$ &  $v_\infty$ & $v({\rm H}\alpha)$ &  Environments$^a$ \\
&    &  &   (\degr)&  (\degr)&   (mag) &  (\km)&  (\km) &  \\ 
\hline
Sk$-$65\degr22	& HD\,270952 & O6\,Iaf+		& 276.40 & --35.75 & 12.07 & 1337 & $270 \pm 15$  & Supergiant shell DEM\,48 (N\,13) \\
Sk$-$67\degr211	& HD\,269810 & O2\,III(f$^*$) 	& 277.70 & --32.16 & 12.28 & 3593 & $285 \pm 17$  & \hii\ region DEM\,241 (N\,59A) \\
Sk$-$69\degr246	& HD\,38282  & WN6h	 	& 279.38 & --31.66 & 11.15 & 1835 &252 	          & X-ray bright superbubble 30\,Dor \\
Sk$-$71\degr45	& HD\,269676 & O4--5\,III(f)	& 281.86 & --32.02 & 11.51 & 2488 & $227\pm 19$   & Superbubble DEM\,221 (N\,206) \\
\hline
\end{tabular}
Note: Spectral types and terminal velocities are from \citet{walborn02} and \citet{massa03} for Sk$-$65\degr22,
Sk$-$67\degr211, and Sk$-$71\degr45 and from \citet{willis04} for Sk$-$69\degr246. 
$v({\rm H}\alpha)$ are from \citet{danforth03}, except towards Sk$-$69\degr246 where it comes
from \citet{chu94} and references therein.
Velocities are in the LSR frame. $a:$  from \citet{chu94}. 
The designations DEM  and N refer to entry in the catalog of \citet{davies76} and \citet{henize56},
respectively.
\end{minipage}
\end{table*}

\subsection{The {\em FUSE}\ observations}

The \fuse\ observations are from the PI team programs P103 (PI: Sembach)
and P117 (PI: Hutchings).  
Descriptions of the \fuse\ instrument  design and  inflight performance
are found in Moos et al. (2000) and Sahnow et al. (2000).  The
LiF\,1A, LiF\,2B, SiC\,1A, and SiC\,2B segment spectra cover the wavelength
region of the \ovi\ $\lambda$$\lambda$1031.926, 1037.617 doublet.  However, we have
chosen not to use the SiC\,2B data because of the strong fixed pattern
noise and relatively low resolution of observations near \ovi\ when
using this segment.  
Observations in the SiC\,1A and LiF\,2B segments have  $\sim$30\% and $\sim$60\%  of
the sensitivity of the LiF\,1A segment, respectively. Therefore, we principally use
the LiF\,1A segment for our measurements. We did not 
find any difference between the \ovi\ profiles observed in the different segments
(within the signal-to-noise). We did not combine the
measurements from the three different segments to maintain 
the optimal spectral resolution.
The observations were obtained in the LWRS (30\arcsec$\times$30\arcsec) apertures.
We use the current calibration pipeline software (CALFUSE version
3.0.8 and higher) to extract and calibrate the
observations.    

The zero point in the final wavelength scale was
established by first estimating an average {\fuse}\ velocity
using the ISM lines of \siii\ $\lambda$1020.699, \ari\ $\lambda$1066.660, 
\feii\ $\lambda$1055.262. We used these lines because (i) they are not 
too saturated to be able to measure the centroids accurately;
(ii) they are present in the LiF\,1A segment  and encompass
the \ovi\ $\lambda$1031.926 interstellar absorption line, minimising 
spurious wavelength shifts; (iii) they
trace the same gas that the ions used in the STIS wavelength range (\sii, \feii, and \niii). 
We then shift the {\fuse}\ wavelength to the absolute LSR frame determined
with the accurate STIS spectra. 
We estimate that the relative {\fuse}\ velocity calibration 
of the \ovi\ is accurate to $\sim$5 \km\ after comparing the shifts 
needed for the velocity centroids \cii, \feii, \siii\ in {\fuse}\ to 
match the LSR velocity of similar ions in STIS E140M.

\subsection{The STIS E140M observations}
The STIS E140M Observations were obtained through the GO program 8662 in Cycle 
9 (PI: J. Lauroesch). This program was built as a snapshot survey to complement
the \fuse\ \ovi\ observations by providing coverage of the other highly ionised species, 
\civ, \nv, and \siiv. The STIS data were obtained with the
E140M intermediate-resolution echelle grating.
The entrance slit was set to $0\farcs2\times 0\farcs2$. 
The spectral resolution is 7 \km\ with a detector pixel size 
of 3.5 \km. Information about STIS and its in-flight performance is given 
by Woodgate et al. (1998), Kimble et al. (1998), and in the STIS Instrument 
Handbook (Proffitt et al. 2002).

Standard  calibration and extraction
procedures were employed using the CALSTIS routine (version 2.2). 
The STIS data reductions provide an excellent wavelength calibration with
a velocity uncertainty of $\sim$1 \km. The heliocentric
velocity was corrected to the dynamical local standard of rest-frame (LSR) (Mihalas \& Binney 1981).

\subsection{Selection of species and velocity structures}
The main goal of this work is to study the properties and origin(s)
of the highly ionised species in the LMC. \ovi\ $\lambda$1031.926, \nv\ $\lambda$$\lambda$1238.821,1242.804, 
\civ\ $\lambda$$\lambda$1548.195,1550.770, and \siiv\ $\lambda$$\lambda$1393.755,1402.770 
are the highly ionised species at hand. The excitation potentials are 113.9 eV, 77.5 eV, 47.9 eV, 
and 33.5 eV for \ovi, \nv, \civ, and \siiv, respectively. Therefore, \ovi\ and \nv\ are better
diagnostics of shock-heated gas than \civ\ and \siiv, the latter being susceptible 
to photoionisation. 

To understand the connection, if any, between the highly ionised species 
and the weakly ionised species, we have to investigate the kinematics
of the ions tracing cool/warm and hot gas.
For tracing the cool/warm gas we use the low-ionisation species
\oi\ $\lambda$1302.169 \siii\ $\lambda$1526.707, \sii\ $\lambda$1250.578, 
and \feiii\ $\lambda$1122.524. \oi\ $\lambda$1302.169 and \siii\ $\lambda$1526.707 are 
strong transitions revealing the weaker absorption components along each sight line, 
while \sii\  $\lambda$1250.578
is a weaker transition that reveals the velocity structure of the 
stronger absorption components. \siii\ and \sii\ trace mostly the warm 
neutral gas (WNM), although a fraction may arise in the warm ionised
medium (WIM).  \oi\ is one of the best tracers of \hi\ because
the ionisation fraction of O is coupled to that of H via resonant charge-exchange
reactions; hence, \oi\ should trace only neutral gas along the line of sight. 
\feiii\ $\lambda$1122.524 is found mostly in the WIM, 
and therefore is an excellent substitute for H$\alpha$ emission measurements. 

In Figs.~\ref{spec1} to \ref{spec4}, we show the observed flux
distribution in the vicinity of the highly ionised species with our adopted
continuum fits (see the next sections 
for a full description of the continuum placement for those species), while in Figs.~\ref{specmodel}
and \ref{specmodel1}, we show the normalised
profiles of the weakly and highly ionised species. The continuum near \siii\ 
is modeled with a straight line. The continuum near \oi\ and \sii\ are fitted with low-order
Legendre polynomials. For the weakly ionised species, the most complicated continuum placement 
is near \feiii\ because of very strong stellar wind features from \siiv\ and \pv. 
Near this transition, very high-order Legendre polynomials ($d\ga 5$) must be used,
and this results in a large uncertainty in the continuum placement 
(see Figs.~\ref{specmodel} and \ref{specmodel1}). However, the velocity centroids 
of the LMC \feiii\ components should not be much affected by the continuum placement,
at least near the core of the absorption line.

This set of lines of sight provides an opportunity to 
compare the highly ionised species in various environments from the Galactic halo to the LMC. 
Insight on the highly ionised species at lower velocity is important in its own right, 
but similarities (or lack thereof) in the high-ion kinematics and ratios in the MW and LMC can 
provide further insight on the origin(s) of the highly ionised species in the LMC.  
In Figs.~\ref{specmodel} and \ref{specmodel1}, absorption for the various 
ions is observed from the MW to the LMC velocities. These 
profiles can be separated naturally in four main components: (1) the MW 
component (roughly $v_{\rm LSR} \la 30$ \km), (2) the intermediate-velocity component
(IVC with roughly $30\la v_{\rm LSR} \la 100$ \km), (3) the high-velocity component
(HVC with roughly $100\la v_{\rm LSR} \la 175$ \km), and (4) the LMC components (roughly  
$v_{\rm LSR} \ga 175$ \km). Here, we follow the definition of IVC and HVC from Wakker (1991).
We will argue that the IVC is 
closely connected to the MW, but the high-velocity component may be an HVC with respect to 
the MW or may have an LMC origin.

\subsection{Measurements of column densities, velocities, and line widths of the highly ionised species}

To estimate the column densities in this work, we use the atomic parameters 
compiled by Morton (2003).

\begin{table}
\begin{minipage}{8.5truecm}
\caption{\large Summary of AOD measurements  \label{t2}}
\begin{tabular}{lcccl}
\hline
Ion   & $v_a$  &   $b_a$ &   $\log N_a$&   $\Delta v$ \\
 & (\km)   &   (\km)  &   (cm$^{-2}$) &   (\km)
\\
\hline
\multicolumn{5}{c}{Sk$-$65\degr22} \\
\ovi\   & $258.9 \pm 1.7$  & $49.6 \pm 1.6$   & $14.14 \pm 0.02$		   &LMC [187,350]\\
        & $128.1\pm 1.0$   & $37.8 \pm 1.2$   & $13.98 \pm 0.02$		   &HVC  [88,187]\\
        & $53.0 \pm 0.5$   & $22.9 \pm 0.6$   & $13.95 \pm 0.02$		   &IVC  [23,88]\\
        & $-3.0 \pm 0.6$   & $25.9 \pm 0.4$   & $13.85 \pm 0.03$		   &MW  [--45,23]\\
\nv\    & \nodata	    & \nodata	      & $< 13.26   $			   &LMC [187,350]\\
        & \nodata	   & \nodata	      & $<13.26 	   $		   &HVC  [88,187]\\
        & \nodata	   & \nodata	      & $<13.20 	   $		   &IVC  [23,88]\\
        & \nodata	   & \nodata	      & $<13.27 	   $		   &MW  [--45,23]\\
\hline
\multicolumn{5}{c}{Sk$-$67\degr211} \\
\ovi\   & $251.2 \pm 2.3$  & $53.9 \pm 2.6$   & $14.20 \pm 0.02$		   &LMC [180,355]\\
        & $129.1\pm 1.6$   & $32.5 \pm 1.3$   & $13.87 \pm 0.02$		   &HVC [100,182]\\
        & $ 58.2\pm 1.1$   & $26.9 \pm 0.8$   & $14.13 \pm 0.02$		   &IVC [30,100] \\
        & $3.8  \pm 1.5$   & $22.2 \pm 2.0$   & $13.97 \pm 0.03$		   &MW  [--43,30]\\
\nv\    & \nodata	   & \nodata	      & $<13.26    $			   &LMC [180,355]\\
        &\nodata	   &\nodata	      & $< 13.12	$		   &HVC [100,182]\\
        &\nodata	   &\nodata	      & $< 13.11	$		   &IVC [30,100] \\
        &\nodata	   &\nodata	      & $< 13.17	$		   &MW  [--43,30]\\	      
\hline
\multicolumn{5}{c}{Sk$-$69\degr246} \\
\ovi\   & $215.7 \pm 2.1$  & $46.3 \pm 3.5$   & $14.36 \pm 0.09$		   &LMC [139,302]\\
        & $103.1 \pm 2.7$  & $15.1 \pm 4.9$   & $13.55 \pm 0.20$		   &HVC [85,139]\\
        & $62.7  \pm 2.2$  & $17.4 \pm 2.4$   & $13.59 \pm 0.10$		   &IVC [41,85]\\
        & $5.5  \pm 1.4$   & $32.2 \pm 2.1$   & $13.96 \pm 0.10$		   &MW [--61,41] \\
\nv\    & \nodata	   & \nodata	      & $< 13.40   $			   &LMC [139,302]\\
       & \nodata	   & \nodata	      & $<13.20    $			   &HVC [85,139] \\
        & \nodata	   & \nodata	      & $<13.26    $			   &IVC [41,85]\\
        & \nodata	   & \nodata	      & $<13.61 	   $		   &MW  [--61,41]\\
\hline
\multicolumn{5}{c}{Sk$-$71\degr45} \\
\ovi\   & $227.3 \pm 1.2$  & $53.4 \pm 1.5$   & $14.37 \pm 0.02$		   &LMC [162,320]\\
        & $143.0 \pm 0.9$  & $32.5 \pm 0.7$   & $14.12 \pm 0.02$		   &HVC [104,183]\\
        & $67.7  \pm 0.8$  & $31.2 \pm 0.6$   & $14.20 \pm 0.02$		   &IVC [26,104]\\
        & $-0.2 \pm 1.4$   & $26.0 \pm 1.5$   & $13.77 \pm 0.02$		   &MW [--51,26] \\
\nv\    & \nodata	   & \nodata	      & $<13.51    $			   &LMC  [162,320] \\
        & \nodata	   & \nodata	      & $<13.16 	   $		   &HVC  [104,183]\\
        & \nodata	   & \nodata	      & $<13.21 	   $		   &IVC  [26,104]\\
        & \nodata	   & \nodata	      & $<13.22 	   $		   &MW  [--51,26]\\
\hline
\end{tabular}
Note: Refer to \S\S\ref{ovimeas},\ref{nvmeas} for more details on these measurements.
\end{minipage}
\end{table}

\begin{table}
\begin{minipage}{8.5truecm}
\caption{\large \civ\ and \siiv\ fit results \label{t3}}
\begin{tabular}{lcccl}
\hline
Ion   & $v$  &   $b$ &   $\log N$&   Cloud \\
 & (\km)   &   (\km)  &   (cm$^{-2}$) &   
\\
\hline
\multicolumn{5}{c}{Sk$-$65\degr22} \\
\civ\   & $271.7 \pm  1.7$  & $23.9 \pm  2.7$   & $13.59 \pm 0.05$	 &LMC  \\
        & $265.4 \pm  0.5$  & $ 6.1 \pm  1.2$   & $13.45 \pm 0.07$	 &LMC (\hii)  \\
        & $105.1 \pm 13.9$  & $24.4 \pm 18.2$   & $13.11 \pm 0.09$	 &HVC \\
        & $ 43.1 \pm  0.6$  & $17.3 \pm  0.9$   & $13.84 \pm 0.02$	 &IVC  \\
        & $-11.0 \pm  1.1$  & $13.2 \pm  1.8$   & $13.35 \pm 0.04$	 &MW  \\
\siiv\  & $274.1 \pm 1.0$   & $24.0 \pm  1.5$   & $13.20 \pm 0.03$	&LMC \\
        & $269.0 \pm 0.3$   & $ 6.7 \pm  0.5$   & $13.25 \pm 0.03$	&LMC (\hii)  \\
        & $135.6 \pm 5.4$   & $41.4 \pm 10.2$   & $12.59 \pm 0.07$	&HVC \\
        & $ 47.3 \pm 0.8$   & $17.9 \pm  1.2$   & $12.96 \pm 0.03$	&IVC  \\
        & $-10.3 \pm 0.4$   & $10.7 \pm  0.8$   & $12.96 \pm 0.02$	&MW  \\
\hline
\multicolumn{5}{c}{Sk$-$67\degr211} \\
\civ\   & $272.2 \pm 0.3$   & $11.4 \pm 1.0$    & $(>)15.94 \pm 0.32$	&LMC (\hii) \\
        & $251.2 \pm 3.4$   & $43.1 \pm 2.9 :$    & $13.80 \pm 0.05:$	&LMC  \\
        & $113.2 \pm 4.9$   & $41.0 \pm 8.3$    & $13.48 \pm 0.07$	&HVC \\
        & $ 47.9 \pm 6.7$   & $22.9 \pm 6.8$    & $13.78 \pm 0.22$	&IVC  \\
        & $ 14.0 \pm 5.9$   & $23.7 \pm 4.3$    & $13.89 \pm 0.16$	&MW  \\
\siiv\  & $276.2 \pm 0.3$   & $10.6 \pm 0.9$	& $(>)15.09 \pm 0.25$	&LMC (\hii)  \\
	& $262.0 \pm 1.5$   & $37.4 \pm 1.5$	& $13.53 \pm 0.03$	&LMC   \\
        & $ 38.7 \pm 1.6$   & $31.7 \pm 2.3$	& $13.29 \pm 0.03$	&IVC  \\
        & $  0.1 \pm 0.7$   & $10.5 \pm 1.2$	& $12.90 \pm 0.05$	&MW  \\
\hline
\multicolumn{5}{c}{Sk$-$69\degr246} \\
\civ\   & $214.1 \pm 0.5$   & $16.7 \pm 1.4$    & $(>)14.28 \pm 0.02$	&LMC (\hii) \\
	& $213.5 \pm 1.8$   & $37.0 \pm 7.1:$    & $13.90 \pm 0.14:$	&LMC  \\
        & $ 98.6 \pm 3.5$   & $20.7 \pm 7.9$    & $13.09 \pm 0.09$	&HVC \\
        & $ 50.2 \pm 1.3$   & $14.5 \pm 2.5$    & $13.44 \pm 0.05$	&IVC  \\
        & $  6.7 \pm 1.0$   & $19.6 \pm 1.7$    & $13.75 \pm 0.03$	&MW  \\
\siiv\  & $218.0 \pm 0.3$   & $12.3 \pm 1.2$    & $(>)14.23 \pm 0.08$	&LMC (\hii) \\
        & $214.5 \pm 0.6$   & $37.0 \pm 7.1$    & $13.72 \pm 0.04$	&LMC   \\
        & $102.4 \pm 2.0$   & $17.6 \pm 3.9$    & $12.52 \pm 0.06$	&HVC \\
        & $ 51.3 \pm 1.0$   & $16.8 \pm 1.9$    & $12.85 \pm 0.03$	&IVC  \\
        & $  3.6 \pm 0.5$   & $16.5 \pm 0.8$    & $13.21 \pm 0.02$	&MW  \\
\hline
\multicolumn{5}{c}{Sk$-$71\degr45} \\
\civ\   & $221.0 \pm 0.3$   & $15.4 \pm 0.5$    & $(>)14.56 \pm 0.03$	&LMC (\hii) \\
        & $128.6       :$   & $25.0 :      $    & $13.35      :$	&HVC  \\
        & $ 61.4 \pm 0.9$   & $27.6 \pm 1.5$    & $13.98 \pm 0.02$	&IVC \\
        & $  1.0 \pm 1.0$   & $13.8 \pm 1.7$    & $13.47 \pm 0.04$	&MW  \\
\siiv\  & $221.2 \pm 0.4$   & $14.8 \pm 0.7$	& $(>)14.18 \pm 0.05$	&LMC (\hii) \\
        & $128.6  :	$   & $25.0 :	   $    & $12.85 \pm 0.08$	&HVC  \\
        & $ 66.3 \pm 1.2$   & $19.8 \pm 2.0$    & $13.34 \pm 0.03$	&IVC \\
        & $ -2.6 \pm 1.2$   & $12.7 \pm 2.1$    & $13.05 \pm 0.05$	&MW  \\
\hline
\end{tabular}
Note: The ``:'' symbol highlights the uncertainty in the profile fitting to 
the component where colon is marked. 
The ``$(>)$'' symbol indicates that the line is possibly uncertain 
due to the effects of strong saturation. See
\S\ref{civmeas} for more details on these measurements.
\end{minipage}
\end{table}

\subsubsection{\ovi\ measurements}\label{ovimeas}

Our \ovi\ measurements are based on an analysis of the 1031.926 \AA\ 
transition. The \ovi\ $\lambda$1037.617 line is always contaminated by \cii, 
\cii*, and H$_2$ lines, and therefore cannot be used. 
The \ovi\ $\lambda$1031.926 absorption in the Galactic component 
can be contaminated by molecular lines from the Milky Way and LMC 
and in principle by \cli\ $\lambda$1031.507. 
However, \cli\ is at $-122$ \km\ and is only strong if the molecular fraction is high. 
We searched  for the presence of \cli\ $\lambda$1004.668 in our four sight lines, 
which has a strength similar to \cli\ $\lambda$1031.507, and found none
at Galactic or LMC velocities. 

Three molecular lines can contaminate \ovi\ $\lambda$1031.926. The HD
6--0 R(0) $\lambda$1031.912 is at $-4$ \km\ relative to \ovi\ in
the restframe. No absorption from HD 3--0 R(0) through 8--0 R(0) at
1066.271, 1054.433, 1042.847, 1021.546, and 1011.457 \AA\ was found in
the spectrum of our target stars. Therefore, HD 6--0 R(0) $\lambda$1031.912
does not contaminate the \ovi\ transition. The H$_2$ lines (6-0) P(3) and R(4)
lines at 1031.191 and 1032.356 \AA\ can also contaminate the \ovi\
absorption.  These are found at $-214$ and $+125$ \km\ relative to the
\ovi\ restframe, respectively, and can arise in both the Milky Way and
LMC (see Howk et al. 2002). The imprints of these lines can be clearly
seen in the spectrum of Sk$-$65\degr22, Sk$-$69\degr246, Sk$-$71\degr45,
but not in the spectrum of Sk$-$67\degr211.  To estimate the
contribution to the \ovi\ profiles from the 6--0 P(3) and R(4) lines
of molecular hydrogen, we fitted Gaussian profiles to a number of
$J=3$ and 4 transitions of H$_2$ with line strengths $\lambda f$
similar to the contaminating 6--0 line transitions.  The available
H$_2$ transitions can be found in Table~3 of Howk et al. (2002). We
have used only H$_2$ lines present in LiF\,1A to minimise any effects
of changes in the shape of the instrumental line spread
function. While saturation effects may be significant, the choice of
lines with similar $\lambda f$ circumvent in great part such problems
(see Howk et al. 2002, Wakker et al. 2003).

\begin{figure}
\includegraphics[width = 7.8truecm]{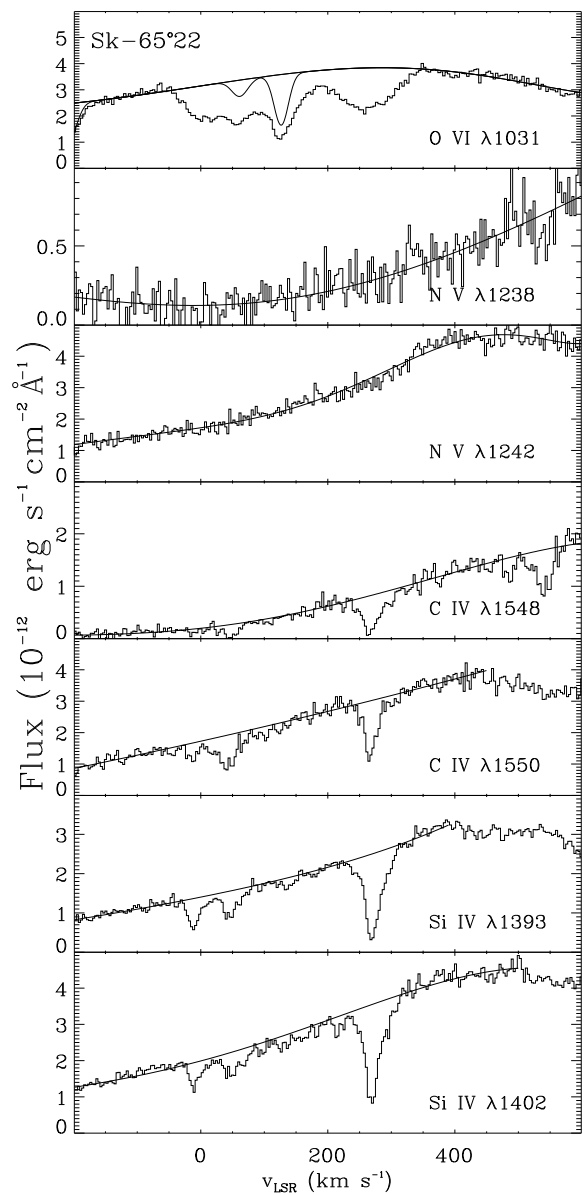}
\caption{Plots of several highly ionised species in flux unit against the LSR velocity towards Sk$-$65\degr22. 
The solid line is the continuum to the interstellar line. In the \ovi\ panel, we also 
the model to the H$_2$ line (6-0) P(3) and R(4)  lines at  1031.191 and 1032.356 \AA.
\label{spec1}}
\end{figure}

A difficulty in assessing the interstellar \ovi\ absorption is that the
\ovi\ stellar component predominantly associated with the wind 
of the star can produce a complicated
continuum. Furthermore stellar wind variability can change the
shape of the continuum and produce discrete absorption components (DACs)
that can mimic \ovi\ interstellar absorption (Lehner et al. 2001).  The
matter is complicated by the fact that we cannot use interstellar
\ovi\ $\lambda$1037.617 (see above).  To evaluate the
effects of the stellar wind, we considered the spectra of \ovi\
$\lambda$$\lambda$1031.926, 1037.617 over a large velocity range in
the \ovi\ vicinity following Lehner et al. (2001).  We note that
there is no observed variation between the different exposures that
were used to produce the final spectra.  We review the continuum
placement and possible contamination from H$_2$ and stellar wind features for each star:

{\em Sk$-$65\degr22:} Its stellar wind has a terminal velocity $v_\infty =-1350$ \km, which is less in absolute value 
than the 1650 \km\ velocity
separation between the lines of the \ovi\ doublet. The interstellar measurements are therefore not strongly affected
by DACs (Lehner et al. 2001). We note that the stellar \ovi\ profiles can be observed predominantly in both lines near $-1000$ \km. 
We are therefore confident that the stellar continuum near \ovi\ $\lambda$1031.926 can be simply
modeled as shown in the top panel of Fig.~\ref{spec1}. We also show in this figure the H$_2$ model, 
where the Galactic component is weaker than the LMC component.

\begin{figure}
\includegraphics[width = 7.8truecm]{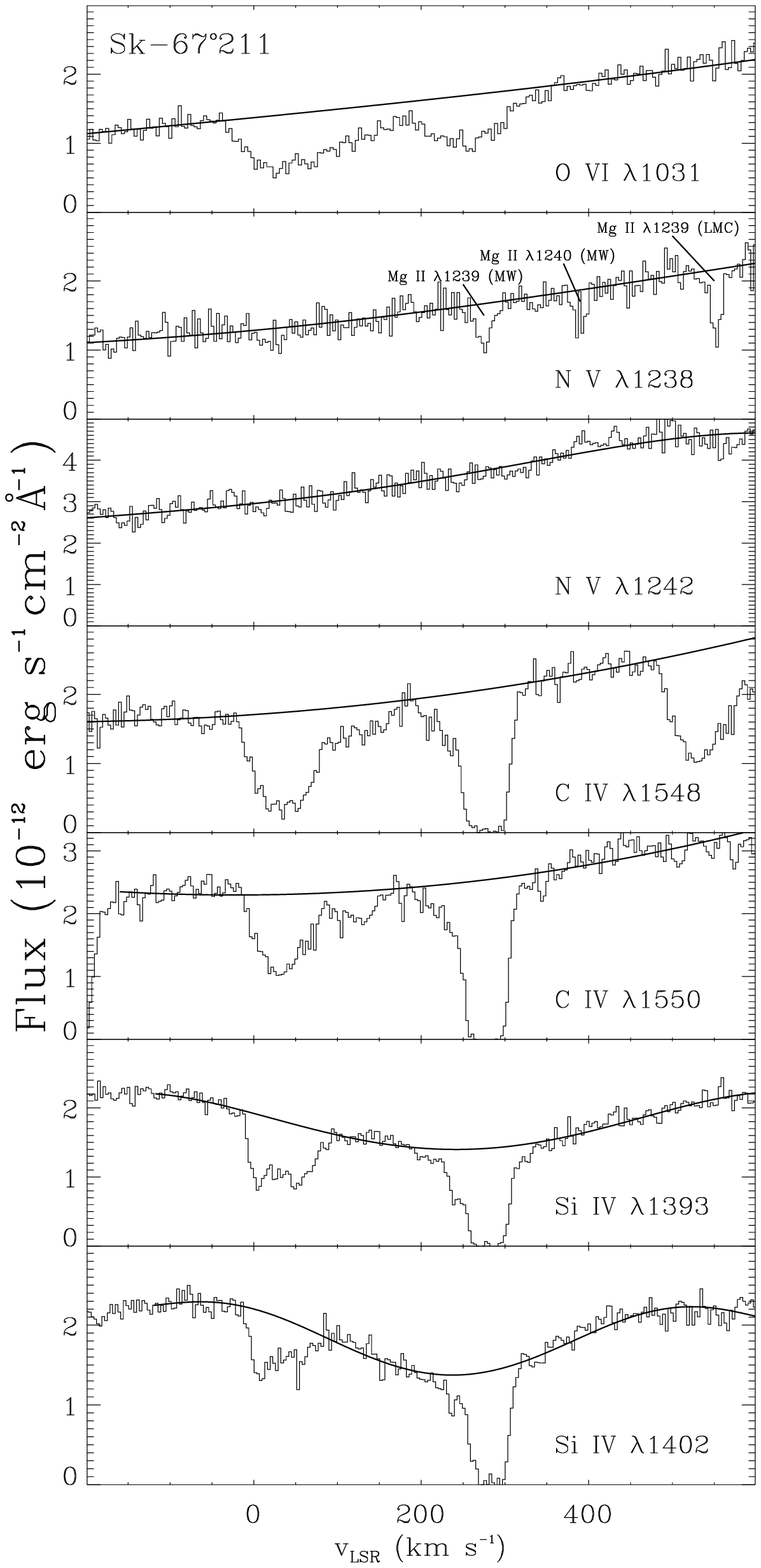}
\caption{Same as Fig.~\ref{spec1} but towards Sk$-$67\degr211. \label{spec2}}
\end{figure}

{\em Sk$-$67\degr211:} This O2 supergiant star has the simplest stellar continuum near
\ovi\ in our sample displaying a very prominent P-Cygni profile with $v_\infty = -3600$ \km\ (see Fig.~\ref{spec2}). 
The effect of DACs on the local continuum is negligible since 
DACs are only strong and narrow near $v_\infty$ (Lehner et al. 2001).  The H$_2$ 
contamination is also insignificant towards this star.  

\begin{figure}
\includegraphics[width = 7.8truecm]{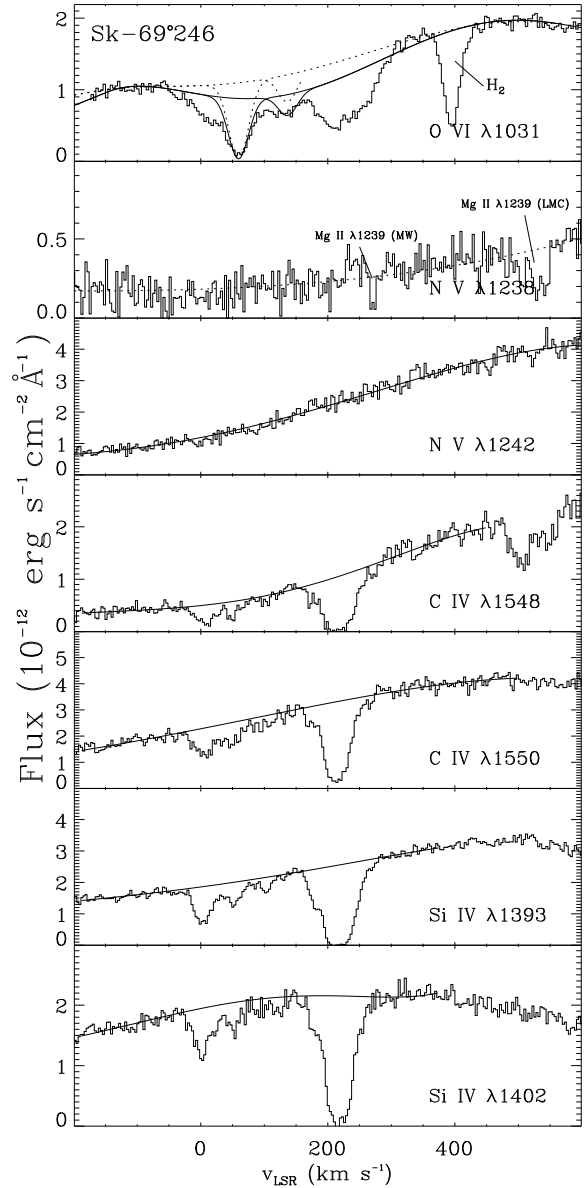}
\caption{Same as Fig.~\ref{spec1} but towards Sk$-$69\degr246 (the dotted line 
shows another possible choice for the continuum). \label{spec3}}
\end{figure}

{\em Sk$-$69\degr246:} This is certainly the most complicated star in
our sample for estimating the interstellar \ovi. At a terminal
velocity of $v_\infty =-1847$ \km, DACs from the \ovi\
$\lambda$1037.617 may be present in the galactic and LMC components of
\ovi\ $\lambda$1031.926.  Moreover, at such terminal velocity, the
wind features from \ovi\ $\lambda$1031.926 are lost in \hi\
Lyman$\beta$, making it impossible to identify features that may be
present in the weaker member of the doublet using the stronger member.
Therefore, it may be possible that the \ovi\ $\lambda$1031.926
interstellar profile is contaminated. We make the
assumption it is not. The continuum placement is 
also uncertain. In Fig.~\ref{spec3}
(top-panel) we show two likely continua. Using the H$_2$ modeling,
the continuum cannot be lower than the solid continuum, otherwise we
would underestimate the amount of the H$_2$ in the LMC component.

\begin{figure}
\includegraphics[width = 7.8truecm]{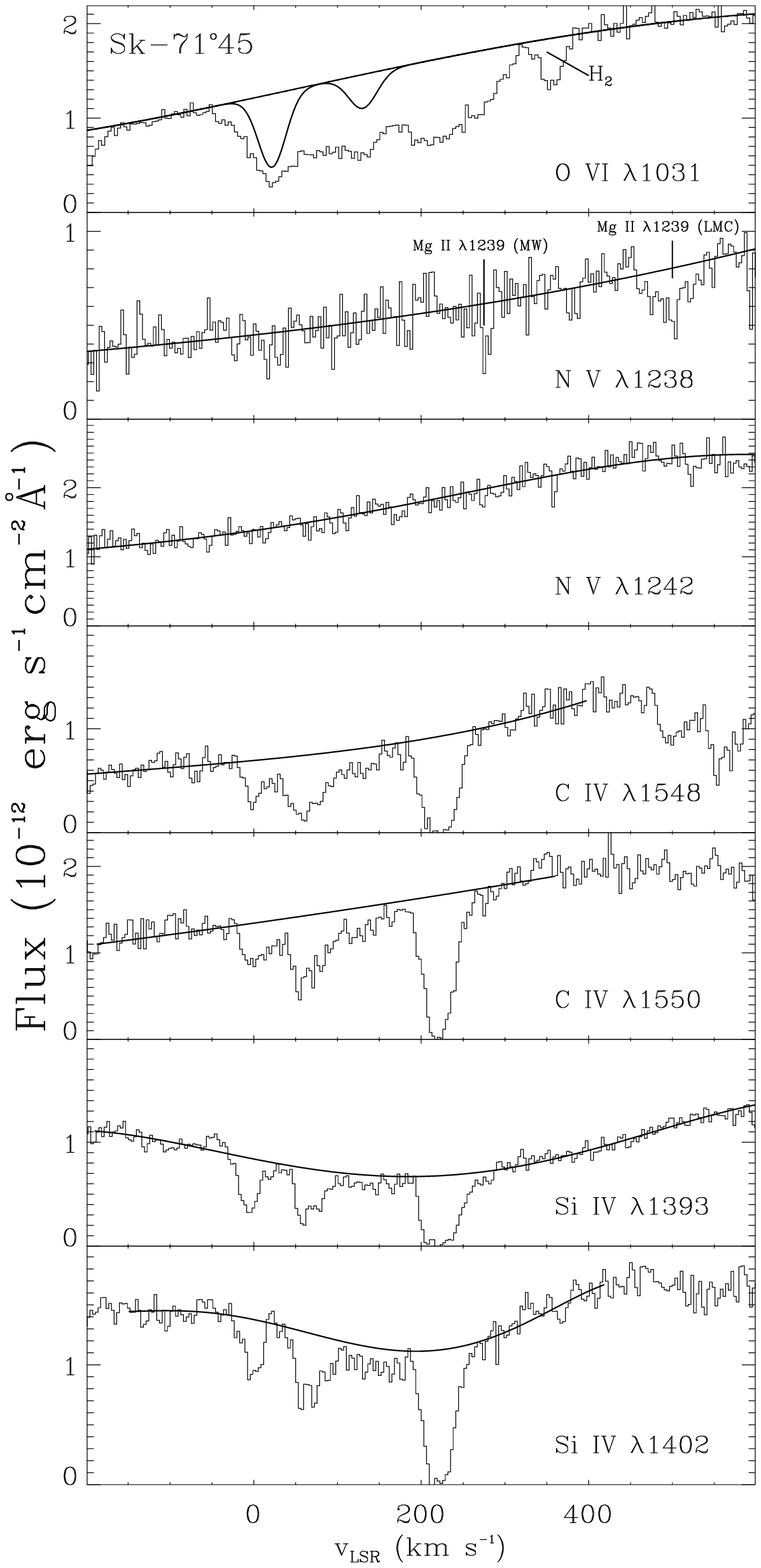}
\caption{Same as Fig.~\ref{spec1} but towards Sk$-$71\degr45. \label{spec4}}
\end{figure}

{\em Sk$-$71\degr45:} This star has a large terminal velocity
($v_\infty = -2500$ \km), and therefore the interstellar \ovi\ profile
is unlikely to be contaminated by a DAC. The continuum is well modeled with
low-order polynomial function (see Fig.~\ref{spec4}). The Galactic
H$_2$ is stronger than the LMC component towards this star.

The \ovi\ absorption does not present distinct components, as can be
observed in the profiles of the other ions. We adopted the
apparent optical depth (AOD) method of Savage \& Sembach (1991) to
estimate the column density ($N_a$), velocities ($v_a$), and line
widths ($b_a$) of \ovi.  The absorption profiles were converted into
apparent column densities per unit velocity $N_a(v) = 3.768\times
10^{14} \ln[F_c/F_{\rm obs}(v)]/(f\lambda)$ cm$^{-2}$\,(\km)$^{-1}$, where $F_c$ is the
continuum flux, $F_{\rm obs}(v)$ is the observed flux as a function of
velocity, $f$ is the oscillator strength of the absorption and
$\lambda$ is in \AA.  The values of $v_a$, $b_a$, $\log N_a$ are
obtained from $v_a= \int v N_a(v)dv / N_a $, $b_a = [2 \int
(v-\bar{v}_a)^2 N_a(v)dv / N_a]^{1/2}$, and $N_a= \int N_a(v)dv$ (see
Sembach \& Savage 1992).

The \ovi\ absorption profiles do not
contain easily distinguished absorbing components, gas
associated with the LMC and gas at lower velocities is usually easily
separated. An important aspect of our investigation involves
studying the kinematical relationships between the high and low 
ions and the column density ratios of the highly ionised species.  We used the \civ\ and \siiv\
profiles to determine velocity integration limits for \ovi\ that would
allow us to crudely separate the different components (MW, IVC,
HVC, and LMC).  

The measurements for \ovi\ are presented in Table~\ref{t2}. For
Sk$-$69\degr246, we adopt the mean results of the high and low
continua. The listed $\pm 1\sigma$ errors are from a quadrature addition of the
statistical error and the continuum placement error. The
$1\sigma$ errors may be slightly underestimated (by no more than 0.05
dex in the log of the column density although usually lower) because they do not include
systematic from the removal of the H$_2$ contamination.
Finally, note that saturation is unlikely to affect our measurements
since the broadening of \ovi\ is greater than the instrumental broadening
of {\em FUSE}.

\subsubsection{\nv\ measurements}\label{nvmeas}
The interstellar \nv\ measurements are the most straightforward because
the strong stellar wind makes the continuum placement simple. 
\mgii\ $\lambda$1239.925 is at 270 \km\ relative to \nv\ $\lambda$1238.810, 
and therefore can contaminate the LMC component of the strong line of the doublet
(see panel 2 in Fig.~\ref{spec2}). 

No interstellar Galactic or LMC \nv\ is detected towards any of the
stars in our sample. The flux is much higher in the weaker transition at
1242.804 \AA\ than in the transition at 1238.810 \AA, and therefore
\nv\ $\lambda$1242.804 produces better upper limits.  We report the
3$\sigma$ upper limit to the column density in Table~\ref{t2} by
assuming the line is on the linear part of the curve of growth.

\subsubsection{\civ\ and \siiv\ measurements}\label{civmeas}

For \civ\ and \siiv, the continuum placement and the measurements of 
$v,N,b$ can be constrained by considering  both lines of the doublet. 
The absorption profiles of these ions appear also to be separated into several 
components rather than the single broad absorption observed in the \ovi\ line profile.
 Furthermore the \civ\ and \siiv\ profiles 
reveal that in most cases the absorption of these ions is coupled, i.e.
both ions show very similar profiles. 
These facts motivated us to use a Voigt profile fitting method. 
The fit results and the errors were obtained using the Voigt component
fitting software of Fitzpatrick \& Spitzer  (1997).  The fit results 
listed are based on a simultaneous fit to both members of the \civ\ doublet on one hand
and the \siiv\ doublet on the other hand. However, we choose not to fit the \civ\ and the \siiv\ 
profiles simultaneously to avoid making an a priori assumption that these two ions 
trace the same gas. Rather, we prefer to show if this is the case (or not) {\em a posteriori}. 
For the STIS E140M instrumental spread function we adopted the 
profile from the STIS Instrument Handbook (Profitt et al. 2002). 
Data taken through the $0\farcs2\times 0\farcs2$ slit have an line spread
function with a narrow core and a broader halo. 

While we do not present the results of the AOD measurements for 
\civ\ and \siiv, we undertook them to cross-check the validity 
of the profile fitting results. We find very similar results between
the AOD and profile fitting methods for the MW and IVC components. 
For the HVC component, the column densities of \civ\ and \siiv\ could be 0.05--0.2 dex
higher with the AOD measurements because  the velocity interval over 
which the profile was integrated was larger than the breadth of the model. 
Hence, the profile fits to the HVC component are uncertain, and the AOD
column density ratios (see \S\ref{aodratiot}) are more reliable since there is no assumption on
the profile of the absorption line. For the strong LMC components,
the AOD provided in general a lower limit in agreement with the profile 
fitting measurements; and in the cases where the LMC components were essentially 
resolved (\civ\ and \siiv\ towards Sk$-$65\degr22 and \civ\ towards Sk$-$69\degr246),
the AOD and profile fitting gave the same results within the 1$\sigma$ errors.

\begin{table}
\begin{minipage}{8.5truecm}
\caption{\large Temperatures and column density ratios \label{t4}}
\begin{tabular}{lcccl}
\hline
Sight line   & $T$  &   $N($\civ$)/N($\siiv) &  $N($\ovi$)/N($\civ) \\
 & ($10^5$ K)   &   &   
\\
\hline
\multicolumn{4}{c}{LMC  -- \hii\ region/Superbubble} \\
Sk$-$65\degr22	& $\le 0.27$	& $1.6\pm 0.3	          $ &  \nodata$^a$		   \\
Sk$-$67\degr211	& $(<)0.2 \pm 0.4 $& $7.1 \pm\,^{9.5}_{4.8}(:)  $ & \nodata$^a$		  \\ 
Sk$-$69\degr246	& $(<)1.6 \pm 0.7 $& $1.1 \pm 0.2	(:)	  $ & \nodata$^a$		  \\ 
Sk$-$71\degr45   & $(<)0.2 \pm 0.3$	& $2.4 \pm 0.3	(:)	  $ & \nodata$^a$		  \\
\hline
\multicolumn{4}{c}{LMC  -- not associated with  \hii\ region}\\
Sk$-$65\degr22	& $<4.2$    	& $2.5 \pm 0.3	       $  & $  3.6 \pm 0.5     $    \\
Sk$-$67\degr211	& $6.0 \pm 3.0 $& $1.9 \pm 0.3:		  $ & $ 2.5 \pm 0.3:    $ \\
Sk$-$69\degr246	& $<10$     	& $1.5 \pm 0.5:		  $ & $ 3.4 \pm 1.4:   $ \\
\hline
\multicolumn{4}{c}{HVC }\\
Sk$-$65\degr22	& $<4.4$	& $3.3 \pm 1.0	      $      & $ 7.4 \pm 1.6   $  \\
Sk$-$67\degr211	& $<12 $	& \nodata  		     & $ 2.5 \pm 0.4   $  \\
Sk$-$69\degr246	& $1.5 :$	& $3.7 \pm 0.9		   $ & $ 2.9 \pm 1.7   $  \\
Sk$-$71\degr45	& \nodata	& $3.2 : \pm 0.7:		   $ & $ 5.9  \pm 0.7	$ \\
\hline
\multicolumn{4}{c}{IVC }\\
Sk$-$65\degr22    & $<2.2$    & $7.6 \pm 0.6		$    & $ 1.3 \pm 0.1   $  \\
Sk$-$67\degr211    & $<3.8 $  & \nodata  		     & $ 2.3 \pm 1.6   $  \\
Sk$-$69\degr246   & $<1.6$    & $3.9 \pm 0.6		   $ & $ 1.4 \pm 0.4   $  \\
Sk$-$71\degr45    & $4.7 \pm 1.4 $    & $4.4 \pm 0.4		   $ & $ 1.7  \pm 0.1	$  \\
\hline
\multicolumn{4}{c}{MW}\\
Sk$-$65\degr22    & $0.8 \pm 0.4 $    & $2.5 \pm 0.3		$    & $ 3.2 \pm 0.4   $  \\
Sk$-$67\degr211    & $ <1.9$   & \nodata		             & $ 1.2 \pm 0.5   $  \\
Sk$-$69\degr246   & $1.5 \pm 0.9 $    & $3.5 \pm 0.3		   $ & $ 1.6 \pm 0.4   $  \\
Sk$-$71\degr45    & $0.4 \pm 0.9$    & $2.6 \pm 0.4		   $ & $ 2.0  \pm 0.2	$  \\
\hline
\end{tabular}
Note: The temperatures are derived from the $b$-values of \civ\ and \siiv.  
The ``:'' symbol indicate some uncertainty in the result that is not 
reflected by the formal errors deduced from the measurements. 
The ``$(<)$'' symbol indicates that these lines are strong and may be composed
of several components. 
The ``$(:)$'' symbol indicates that these lines are saturated and 
therefore the ratio is more uncertain than the formal errors from the profile
fitting suggest. $a$: No \ovi\ is found to associated with the narrow \civ\ 
component. 
See also \S\ref{aodratiot} for the high ratio measurements. 
\end{minipage}
\end{table}

Below, we provide further insights on the choices of the continua and component
fitting  for each line of sight. 

{\em Sk$-$65\degr22:}  The \civ\ stellar wind is really strong, reducing subsequently
the flux near the MW component of the \civ\ $\lambda$1548.195 (see Figs.~\ref{spec1}
and \ref{specmodel}). For \siiv, the chosen continuum is somewhat intermediate 
between plausible slightly higher and lower continua, but within the $1\sigma$ errors
the results can be reproduced by either solution. 
The MW, IVC, and HVC absorption can each be fitted well with one component.
The HVC kinematical structure remains largely uncertain 
and unknown; there is almost certainly more than one
component over the velocity range $[100,180]$ \km. The LMC absorption 
is fitted well with a broad and a narrow component.

{\em Sk$-$67\degr211:} Along this line of sight, the continuum to
\siiv\ is uncertain, especially for the HVC component. Since there is
a \civ\ HVC component, the continua to the \siiv\ lines are likely to
be somewhat higher than seen in Fig.~\ref{spec2} in the HVC range, but there is no useful
constraint to indicate how much higher. Our choice of continua
therefore places a firm lower limit on the \siiv\ column density of
this HVC.  A higher continuum for \siiv\ would also likely change
slightly the IVC results.  In contrast, the continuum to \civ\ is more
straightforward.  This is the only line of sight in our sample where
\siiv\ and \civ\ in the MW and IVC components may trace
different plasmas since the profile shapes and kinematics of \siiv\ and \civ\ are
not consistent with each other. For the LMC, the \civ\ can be fitted
well with a broad and a narrow component. For LMC \siiv\ absorption, the
fit is not unique: one can fit the LMC absorption with a narrow and a
broad component or with three narrow components. We favor the two-component 
fit because {\em a posteriori}\ the kinematics match well the
kinematics derived for \civ. This produces further uncertainties on
the physical parameters derived for \siiv, and we highlight the
uncertainty in Table~\ref{t3} by a colon (we discuss in \S\ref{sec-kin}
that a two-cloud model for the narrow \civ\ and \siiv\ components is likely). 

{\em Sk$-$69\degr246:} While for \ovi\, this line of sight has the most complicated  continuum placement 
near the interstellar \ovi, for \civ\ and \siiv\ the continuum placement is comparatively
straightforward. For the MW, IVC, and HVC, we again use a single 
cloud model for each of these components. The \civ\ $\lambda$1548.195 may 
suggest a two-component structure at about 3 and 25 \km,  and a narrower component for the 
IVC, but within the S/N and considering the other transition, we did not manage to improve statistically the 
fit with such a model. For the LMC, we used again two components. Yet a combination of two narrow or narrow
and broad components can reproduce well the observed profiles. For \civ, the reduced-$\chi^2$
are essentially the same, 1.00 and 1.03 for the narrow-narrow fit and narrow-broad fit, respectively. 
For \siiv, the reduced-$\chi^2$ is slightly better for the narrow-broad component fit compared to the
narrow-narrow component fit, 1.12 compared to 1.27, respectively. 
The centroid of the broad component of \siiv\ and \civ\ matches well the centroid of the
broad \ovi\ absorption. The broadenings of \ovi\ and \civ\ also appear compatible within
$1\sigma$. Finally, the extra absorption observed in the red wing of \civ\ 
and \siiv\ can  only be reproduced adequately with a broad component, putting
more weight on this solution. We therefore favor the broad-narrow component fit, but 
we highlight the uncertainty in the broad component in Table~\ref{t3} by a colon. 
We note that the LMC absorption lines are very strong; if a narrower component
is present in the profile, the column densities could be uncertain by an order of magnitude. 

{\em Sk$-$71\degr45:} This is the second giant star having a similar stellar 
spectrum near \siiv\ to that of Sk$-$67\degr211. Although we had less
difficulty finding continua that appear satisfactory 
for both lines of the \siiv\ doublet than for Sk$-$67\degr211, the true shape of the continua remains unknown 
and the quoted errors do not adequately reflect this systematic uncertainty. As for the other lines of sight, 
the continua for the \civ\ doublet lines are comparatively straightforward to model. The 
\civ\ and \siiv\ absorption profiles are modeled with a single component for each cloud. 
For the HVC, we fixed the $b$-value for both ions; the centroid velocity for
\siiv\ was also fixed to the \civ\ velocity. If $b$ is not fixed, the fitting procedure
tends to fit an unrealistically broad component. The LMC component of 
\siiv\ $\lambda$1393.755 appears to have a wing not reproduced by the fit. However, 
this is not observed in the other transition, and we attribute this to the continuum
placement. The addition of another LMC component to model this absorption  fails to produce
an adequate fit to the lines. The LMC lines are very strong and if a narrower component
is present in the profile (and it is likely the case, see discussion in \S\ref{sec-kin}), the column densities could be again uncertain by an order of magnitude.


\subsection{Column density ratios of the highly ionised species}\label{aodratiot}

\begin{figure*}
\includegraphics[width = 14truecm]{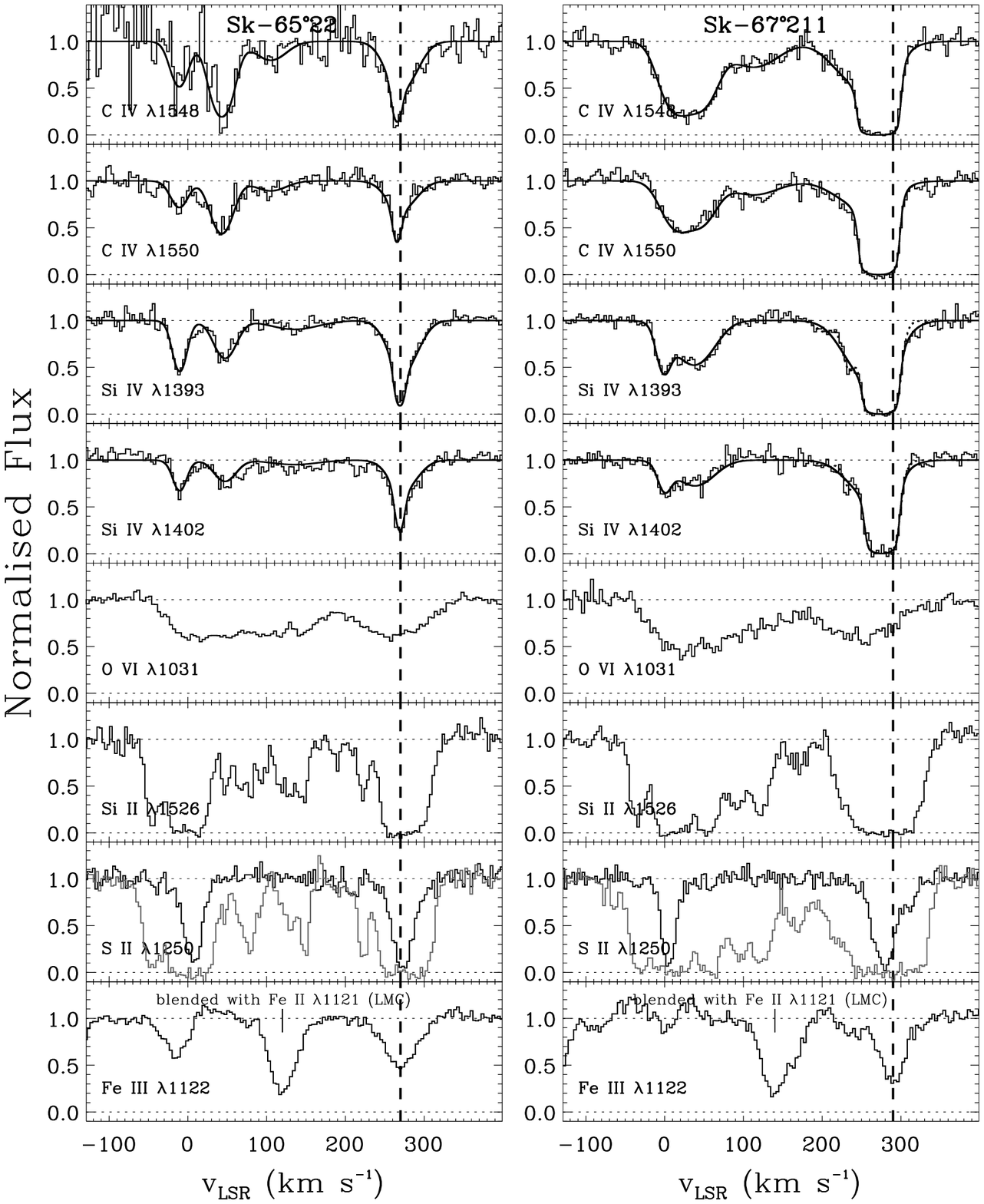}
\caption{Normalised profiles of the selected ions observed
along each line of sight. 
In the \sii\ panel, the gray histogram represents the \oi\ $\lambda$1302 absorption
profile. 
The solid black curve shows the profile fit to the 
\civ\ and \siiv\ lines. We note that the continuum placement near \feiii\ is typically
 uncertain, and these profiles are shown to demonstrate only the velocity distribution
of the \feiii\ absorption. The vertical dashed line represents the systemic H$\alpha$ velocity. 
\label{specmodel}}
\end{figure*}

\begin{figure*}
\includegraphics[width = 14truecm]{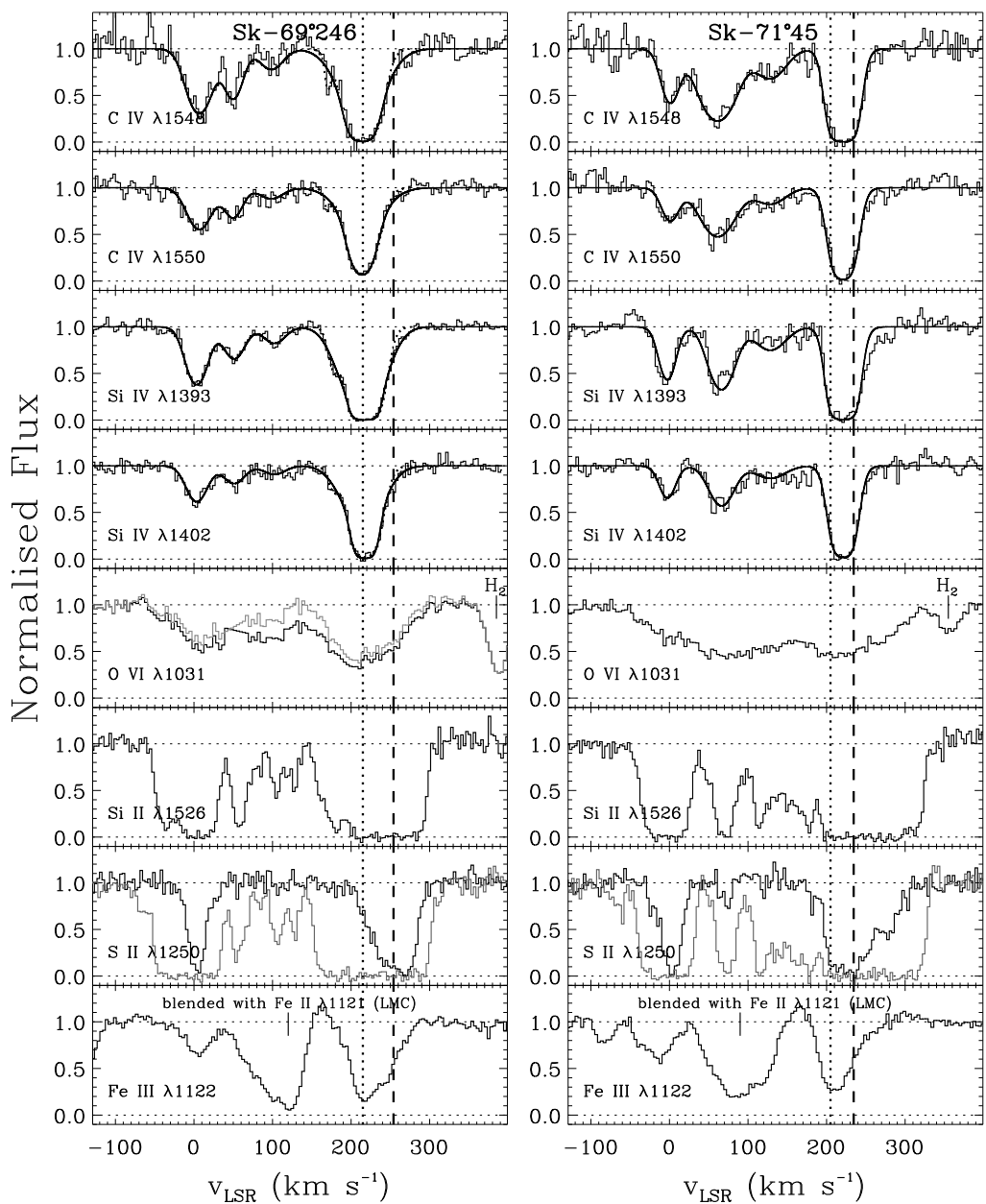}
\caption{Same as Fig.~\ref{specmodel} for the two  sight lines Sk$-$69\degr246 and Sk$-$71\degr45.  
Towards Sk$-$69\degr246, we show the normalised profiles of \ovi\ from both
the high and low continua).  The \sii\ panels also include the \oi\  $\lambda$1302 profiles.
The vertical dotted line represents the expanding H$\alpha$ shell velocity (Y.-H. Chu 2006, priv. comm.). 
\label{specmodel1}}
\end{figure*}

\begin{table*}
\begin{center}
\begin{minipage}{12truecm}
\caption{Ratio of the highly ionised species from AOD comparison \label{t5}}
\begin{tabular}{lccc}
\hline
Model/Sight line   & $N($\ovi$)/N($\siiv)  &   $N($\ovi$)/N($\civ) &  $N($\ovi$)/N($\nv) \\
\hline
\multicolumn{4}{c}{Models} \\
Collisional Ionisation Eq. (CIE)& $\ge 111$	& 3.7--200	& 0.65--55.6     	  \\
Radiative Cooling (RC)		& $\ge 76.9$	& 5.9--27.8	& 11.1--18.9     	  \\
Conductive Interface (CI)	& $\ge 17.2$	& 1.1--23.8	& 2.0--15.4     	  \\
Turbulent Mixing Layers (TMLs)	& 0.8--11.5  	& 0.06--1.0	& 1.7--5.9       	  \\
Shock Ionisation (SI)		& $\ge5.6$  	& 0.96--62.5	& 19.6--32.3      \\
Supernova Remnant (SNR)		& $\ge 74.4$  	& 5.2--8.4	& 11.9--16.0      \\
Stellar Wind (SW)		& 100	  	& 6.3		& 15.8      \\
Halo SNR    			& \nodata  	& 0.42--5.0	& 3.0--11.1	\\
\hline
\multicolumn{4}{c}{LMC  -- \hii\ region} \\
Sk$-$65\degr22   	       & $ \la 1.0      $& $\la 0.8 $	       &  \nodata 		     \\
Sk$-$67\degr211		       & $< 0.3        $& $< 0.1$	      &  \nodata   \\	    
Sk$-$69\degr246		       & $< 0.8  $	& $< 0.4  $	      &  \nodata  	    \\
Sk$-$71\degr45		       & $< 0.5  $	 & $\la 0.2$	      &  \nodata     \\
\hline
\multicolumn{4}{c}{LMC  -- not associated with  \hii\ region} \\
Sk$-$65\degr22  [187,232]       & $>37$          & $>5.4$	       & $>3 $	   	     \\
Sk$-$67\degr211 [180,230]       & \nodata        & 2.2--5.1  	       & $>7  $	   	  \\
Sk$-$69\degr246 [139,190]       & 4.6--10.7	& 1.7--5.0	       & $>10  $   	\\
Sk$-$71\degr45	 [162,190]     &  \nodata  	& 5.4--9.1	       & $>9$          \\
Sk$-$71\degr45	 [260,320]     &   \nodata	& $>11.0$	       & $>6$      \\
\hline
\multicolumn{4}{c}{HVC  } \\
Sk$-$65\degr22   [88,187]       &  20.4--57.5    & 4.7--9.5	       & $>5 $	   	  \\
Sk$-$67\degr211	 [100,182]     & \nodata  	& 2.9--4.3	       & $>6 $      \\
Sk$-$69\degr246	 [85,139]      &  8.1--30.1    	& 1.4--5.6	       & $>2$	       \\
Sk$-$71\degr45	 [104,183]     &  \nodata    	& 4.9--7.2	       & $>9$      \\
\hline
\multicolumn{4}{c}{IVC  } \\
Sk$-$65\degr22   [23,88]        &   4.0--12.6   & 0.4--1.3	       & $>6 $	   	  \\
Sk$-$67\degr211	 [30,100]      &   7.9--20.1   & 0.9--2.2	       & $>11 $   \\
Sk$-$69\degr246	 [41,85]       &   3.9--7.9    & 0.9--1.7	       & $>2  $    	 \\
Sk$-$71\degr45	[26,104]       &   3.2--20.0   & 1.0--3.2	       & $>10 $    	\\
\hline
\multicolumn{4}{c}{MW } \\
Sk$-$65\degr22   [--45,23]      &   2.5--11.2   & 1.1--3.2	       & $>4 $	   	 \\
Sk$-$67\degr211	[--43,30]      &   4.0--7.1    & 1.0--1.8	       & $>6 $	   \\
Sk$-$69\degr246	[--61,41]      &   3.2--6.3    & 1.1--2.0	       & $>2   $   	 \\
Sk$-$71\degr45	[--51,26]      &   2.5--7.1    & 1.0--1.8	       & $>4  $    	 \\
\hline
\end{tabular}
Note: See \S\ref{aodratiot} for the high ratio measurements. 
References for the models: CIE -- \citet{sutherland93} for gas temperature $T= (2$--$5)\times 10^5$ K; RC -- 
\citet{edgar86}; CI: \citet{borkowski90} for magnetic field orientations $\theta = 0$--85\degr and 
interface gas ages $10^{5-7}$ yr; TML -- \citet{slavin93} for a gas flow in the range
25--100 \km\ and temperature of postmixed gas in the range $\log T = 5$--5.5; SI -- \citet{dopita96}
for $150 < v_{\rm shock} < 500$ \km\ and a magnetic parameter less 4 $\mu$G\,cm$^{-3/2}$;
SNR -- \citet{slavin92} for SNR ages  $10^{5.6-6.7}$ yr; SW -- \citet{weaver77}; 
Halo SNR --\citet{shelton98} for SNR ages $10^{6.0-7.2}$ yr.
\end{minipage}
\end{center}
\end{table*}

Theoretical models predict the highly ionised species are produced via a range of
mechanisms, such as in shocks or in interfaces 
between different temperature gas, with conductive heating, turbulent
mixing, or in radiative cooling gas (see references in Table~\ref{t5} and summaries
from Spitzer 1996, Fox et al. 2003 Indebetouw \& Shull 2004). \ovi\ is of
prime importance since in the galactic environments, it is produced
mainly via collisional processes. In contrast \siiv\  and
\civ\ can be produced by photoionisation. If there is 
a kinematical relationship between the highly ionised species, the ratio of the 
high-ion column densities can be used to determine which physical
mechanism is predominant in the ISM, if any, and how the
mechanisms responsible for producing the highly ionised species may change in different galactic environments.

Two methods were used to determine the ratios of highly ionised species. In
Table~\ref{t4}, the ratios of the column densities of \civ\ and \siiv\
obtained from the profile fitting are summarised. We also measured
systematically the \ovi\ column densities and \nv\ column density
limits by using the velocity extent of each cloud defined by the profile
fitting of \civ\ and \siiv.  The ratios $N($\ovi$)/N($\civ) derived
from these measurements are listed in the last column of Table~\ref{t4}. Since the \ovi\
absorption profiles do not break up into easily-identifiable clouds or
blends of clouds like \civ\ and \siiv, \ovi\ may or may not be
associated with the other ions.

In the second method, we compare the apparent column density profiles
over a certain velocity range that is listed in Table~\ref{t5}. This method does
not make any physical assumption of the properties of the gas. We
separate the profiles in several velocity ranges corresponding to
various clouds obtained in the profile fitting (MW, IVC, HVC,
LMC). We give in Table~\ref{t5} the range of ratios that
spans the full range of the $\pm 1\sigma$ errors. If only an upper limit is
available, the $3\sigma$ upper limit and the apparent column density
were estimated over the same velocity range. 

There is a good agreement between both methods for the MW and IVC
components.  This is not surprising since the MW and IVC components
are strong, but not saturated, and well separated from one another and
from other component blends. The component fits to the \civ\ and \siiv\ HVC profiles
are not always unique due to the broad wing-like shape of the HVC
absorption. Yet, there is a rough agreement between the AOD and profile
fitting methods. Because the HVC profiles are not Gaussian, we favor the
AOD results over this velocity range.  
Hereafter, we discuss separately the quite narrow components of the highly ionised species
in the LMC and the broad LMC components, which we assume includes all of the 
\ovi\ given the lack of narrow absorption in this ion.

For the LMC broad component, the ratios derived from the AOD and
profile fitting methods can generally not be compared:  
the AOD estimates the ratios in the wing of the
absorption where it is not contaminated by the narrow components
observed in \civ\ and \siiv, while the profile fitting models
the broad absorption even in regions where the strong narrow absorption 
is present (see Figs.~\ref{specmodel} and \ref{specmodel1} ).  In two cases, the
comparison between the two methods can be made because the AOD estimates 
are made over a large fraction of the profile: Sk$-$67\degr211
 ($v_{\rm LSR} = [180,230]$ \km) and Sk$-$69\degr246  ($v_{\rm LSR} =
[139,190]$ \km).  The AOD \ovi/\civ\ ratios are in excellent agreement with
those derived with the profile fitting for these sight lines. 
For the LMC narrow component, the
only line of sight for which the ratios appear secure is Sk$-$65\degr22;
along this sight line the narrow component of \civ\ and \siiv\ is not too strong
and, therefore, the profiles can be modeled without too much
uncertainty. For the other lines of sight, the high-ion ratios of the
narrow component are uncertain because \civ\ and \siiv\ are saturated.
We discuss the high-ion ratios and implications in \S4. 

\section{Overview of the high-ion kinematic properties}\label{sec-kin}

\begin{table*} 
\begin{minipage}{14 truecm}
\caption{\large Summary of the velocity centroids in the LMC \label{t6}}
\begin{tabular}{lcccccl}
\hline
Name    &    $v({\rm H}\alpha)$ & $v($\feiii)& $v($\ovi)& $v($\civ) (narrow) & $v($\civ) (broad) &  Environments$^a$ \\
&    (\km)&  (\km)&  (\km)&  (\km)&  (\km) &  \\ 
\hline
Sk$-$65\degr22	    &  $270 \pm 15$  & $270.6 \pm 1.3$ & $258.9 \pm 1.7$ & $265.4 \pm 0.5$ & $271.7 \pm  1.7$  & Faint \hii\ region \\
Sk$-$67\degr211	    &  $285 \pm 17$  & $286.3 \pm 1.0$ & $251.2 \pm 2.3$ & $272.2 \pm 0.3$ & $251.2 \pm 3.4$   & \hii\ region\\
Sk$-$69\degr246	    &  252	     & $214.7 \pm 3.1^a$ & $215.7 \pm 2.1$ & $214.1 \pm 0.5$ & $213.5 \pm 1.8$ & Superbubble (X) \\
Sk$-$71\degr45	    &  $227\pm 19$   & $214.8 \pm 1.8$ & $227.3 \pm 1.2$ & $221.0 \pm 0.3$ & \nodata	       & Superbubble \\
\hline
\end{tabular}
Note:  $a$: For this sight line, \feiii\ was fitted with a two-Gaussian component that yields for
the second component $v_{\rm LSR} = 242.3 \pm 4.4$ \km. 
\end{minipage}
\end{table*}

\subsection{Kinematics of the LMC absorption profiles}
Interstellar absorption from the low- and high-ionisation species is
always detected over the full range of velocities from the Milky Way to the LMC 
(except for \nv) along the four lines of sight in our sample.  In Figs.~\ref{specmodel} and
\ref{specmodel1}, we show the normalised profiles of the various
species.  In these figures, the vertical dashed line represents the profile-weighted
average velocities measured for the H$\alpha$ profiles (results from
Danforth 2006 and Chu et al. 1994, see Table~\ref{t6}). Both superbubbles
show evidence for expanding shells in H$\alpha$; the projected radial velocities of the 
approaching side are shown by the dotted lines in Fig.~\ref{specmodel1} 
(Y.-H. Chu 2006, priv. comm.; see also Chu \& Kennicutt 1994). 
There is no obvious structure of expanding shell seen in H$\alpha$ toward the 
other sight lines. The average velocities obtained from the H$\alpha$ emission profiles 
agree remarkably well with the centroids measured in the \feiii\ profiles (see 
Table~\ref{t6}). The \feiii\ profiles in the spectra of
Sk$-$69\degr246 and Sk$-$71\degr45 show indication of two
components with one matching the systemic velocity of the H$\alpha$
profile and the other matching the velocity of the expansion of
the shell. The agreement between the H$\alpha$ and \feiii\ profiles 
is quite remarkable given that H$\alpha$ emission data probe gas in
front of and behind the star while absorption spectra only probe gas
in front the star.

Figs.~\ref{specmodel} and \ref{specmodel1} show continuous absorption
at LMC velocities ($\sim$130--310 \km\ towards Sk$-$65\degr22;
$\sim$210--370 \km\ towards Sk$-$67\degr211; $\sim$130--310 \km\ towards
Sk$-$69\degr246; $\sim$150--340 \km\ towards Sk$-$71\degr45) for \ovi,
\siii, and \oi. (We note that velocities $80\mbox{ \km} \la v_{\rm LSR} \la
130\mbox{ \km}$ may have an LMC origin as well; see \S4.2.)  Over
these velocities, the \ovi\ profiles are broad and have vaguely
Gaussian shapes. The weakly ionised species consist of several narrow
components. Since \ovi\ traces much hotter gas than \siii\ or \oi, the difference 
in the appearance of the profiles is expected. In contrast with the weakly ionised species, 
the overall kinematic morphology of the \ovi\ profiles is quite similar
between the various sight lines.

\siii\ and \oi\ also reveal the presence of an IVC with respect to the LMC
at $v_{\rm LSR} \sim 150$--220 \km, clearly separated from the main absorption 
by roughly $-50$ \km. This LMC-IVC component possibly traces some ejecta from the
LMC disk (see \S4.2).  The \ovi\ absorption is always present in this
IVC component.  We note here that the important difference between the
choice of strong transitions compared to the weaker \feii\ absorption
profiles used by \citet{howk02}: the strong transitions reveal cooler, neutral and partially ionised
gas traced by weakly ionised species and hot gas traced by \ovi\ are systematically
found at similar velocities.

The morphology of the \civ\ and \siiv\ profiles differs from 
the morphology of the \ovi\ profiles. Along our four sight lines, 
a strong narrow component is observed in \civ\ and \siiv\ but not revealed in \ovi. 
Along three sight lines, the profiles 
of \civ\ and \siiv\ have the strong narrow component blended with a broad component 
(see  \S\ref{civmeas}). The broad component is likely associated
with some of the observed \ovi.  There is a good agreement between the
centroids of broad \civ\ and \siiv\ components with the centroid of the broad \ovi\ 
absorption towards Sk$-$67\degr211 and Sk$-$69\degr246 (see summary in
Table~\ref{t6}). Towards Sk$-$65\degr22 the broad component of \civ\ and
\siiv\ is shifted by 12 \km\ because the lower velocities observed 
in the \ovi\ profile are not seen in the \civ\ and \siiv\ profiles.

In the faint \hii\ region where Sk$-$65\degr22 is located, the centroids of the narrow 
components of \civ\ and \siiv\ align with those of \feiii\ and H$\alpha$
as well as with the main component of the weaker ions (see
Fig.~\ref{specmodel}).  This sight line has the simplest LMC kinematics of our four.
For Sk$-$67\degr211, the narrow components of
the \civ\ and \siiv\ absorption profiles appear shifted to lower
velocities relative to the velocities of \feiii\ and
H$\alpha$. However, \civ\ and \siiv\ are saturated and the narrow
profiles may contain more than one component. Towards 
Sk$-$69\degr246, the main absorption of \civ\ and \siiv\ is observed
at the expanding velocity seen in H$\alpha$ and \feiii. There is  
very little absorption in the \civ\ and \siiv\ profiles at 252
\km, which corresponds to the systemic velocity of H$\alpha$.
Finally, towards Sk$-$71\degr45, also located in a superbubble,
the \civ\ and \siiv\ profiles do not align with either component 
of \feiii, their velocities being somewhat intermediate between
the systemic and expanding shell velocities observed in \feiii\ 
and H$\alpha$. The gas traced by  the narrow components of \civ\ and \siiv, 
and \feiii\ are kinematically related in a different manner toward each sight line: 
there is an excellent agreement in their velocity profiles in the quiet 
\hii\ region; in the bright \hii\ region, \civ\ and \siiv\ 
is blue-shifted with respect to \feiii, possibly revealing an highly-ionised 
expanding shell; in the X-ray bright superbubble, \civ\ and \siiv\ appear to probe
mostly the expanding shell seen in the H$\alpha$ profile; and in the 
superbubble DEM\,221, \civ\ and \siiv\ do not appear to directly be kinematically
related to H$\alpha$, with the velocities of the highly ionised species
intermediate between the expanding or systemic velocities of H$\alpha$. We finally note that 
weaker ions (\siii\ and \oi\ for example) and higher ions (\ovi) can produce
absorption at these velocities.

In summary, the absorption profiles of the weakly ionised species 
and \ovi\ always span the same velocity range, though the former are 
always found in discrete, narrow components, while
the latter is broad and widespread. 
There is either a positive, negative, or no shift between the average 
velocities of \ovi\ and weakly ionised species. However, the comparison of the average velocities 
of \ovi\ with the weakly ionised species is complicated since \ovi\ is almost certainly a blend of 
several components (see \S3.2).  
The velocity extent of the \civ\ and \siiv\ LMC profiles is sometimes smaller than the 
\ovi\ profile extent. The narrow LMC components of \civ\ and \siiv\ may be kinematically
connected to the gas traced by H$\alpha$ and the weakly ionised species (e.g., \feiii, \siii), 
but not in a systematic way, implying that even though the narrow components of
\civ\ and \siiv\ probe cool gas ($T\la2\times 10^4$ K, see below) as \feiii, 
their origin/production may differ, at least in some cases. 

\subsection{Broadening of the high-ion absorption profiles}
The gas temperatures derived from the $b$-values of \civ\ and \siiv\ are 
given in Table~\ref{t3}. These remain quite uncertain for the LMC narrow component as the profiles
are very strong, and the errors do not reflect the possibility that the profiles have more than 
one component. To derive these temperatures, we assume that \civ\ and \siiv\ reside in
the same gas, which is supported by the compatibility in the centroids 
of \civ\ and \siiv\ for most clouds. 
In Table~\ref{t3}, we list the temperature of the gas derived from the observed broadening 
of the absorption lines. If $b($\civ$)>b($\siiv) then $T$ can be determined from 
$b = \sqrt{2kT/(Am_{\rm H}) + b^2_{\rm nt}}$, where $b_{\rm nt}$ is the non-thermal component
and the other symbols have their usual meaning.
If $b($\civ$)<b($\siiv) an upper limit on the temperature can only be derived
because we cannot estimate the non-thermal broadening.   

For the strong, narrow LMC component (denoted LMC (\hii) in Table~\ref{t3}), 
we find a good agreement between the centroids of \siiv\ and \civ. 
The Doppler broadenings suggest a temperature of $2\times 10^4$ K along three lines of sight 
(towards Sk$-$69\degr246, within  $2\sigma $ error, the temperature is compatible with 
2--$9 \times 10^4$ K). Since the profiles are strong and saturated towards at least
Sk$-$67\degr211 and Sk$-$69\degr246, the derived $b$ is uncertain but places a firm
upper limit on the temperatures; these clouds likely have
$T \la 2\times 10^4$ K. The non-thermal broadenings of \civ\ and \siiv\ appear to be as large
as the thermal broadening  if the profiles consist of a single component.  

There must be little or no \ovi\ associated with the narrow and strong \civ\ and 
\siiv\ components since there is no indication of a narrow component 
in the overall broad and smooth profile of \ovi\ (see the limits of \ovi\ to \civ\ 
or \siiv\ ratios in Table~\ref{t5}). Hence there is no evidence for \ovi\ 
arising directly within the superbubble shells or the \hii\ regions. 
We  also note that these \hii\ regions and superbubbles 
do not produce any detectable \nv. This is not surprising since the temperatures deduced from
the \civ\ and \siiv\ line-width are only a few $\times 10^4$ K or less, too cool 
to produce any significant amount of \ovi\ or \nv.\footnote{N could be underabundant
in the LMC \hii\ regions (Garnet 1999, Welty et al. 1999). However, the ratios of \civ/\nv\ 
for the narrow component are  greater than 1.7, 478, 7.6, 10.0 towards Sk$-$65\degr22, Sk$-$67\degr211,
Sk$-$69\degr249, and Sk$-$71\degr45, respectively. So even if N may be intrinsically
underabundant by a factor up to 6, with such high ratio limits (especially towards Sk$-$67\degr211),
the bright \hii\ region and superbubbles do not produce any observable amount of \nv. \label{foot-nv}}

For the broad \civ\ and \siiv\ components of the LMC, their $b$-values
imply temperatures of a few $\times 10^5$ K or less. Their broadening is 
consistent with that of \ovi, and these components likely
probe hot gas. For \ovi\ only upper limits to the temperatures can
be derived, and these are always consistent with temperatures of a
few  $10^5$ K. The broad LMC components of the highly ionised species have 
$b({\rm LMC}) \ge \sqrt{b^2({\rm MW})+b^2({\rm IVC})}$, 
suggesting that they likely trace more than two different clouds.

\section{The highly ionised plasma in the LMC}\label{origint}

We found that the absorption of the highly ionised species at velocities similar to the LMC 
disk is dominated by a narrow absorption observed in \civ\ and \siiv\ but not in the
\ovi. The \ovi\ along all four sight lines is  broadly distributed, and 
the profile fitting of \civ\ and \siiv\ reveals broad absorption 
in these ions along 3 of 4 sight lines.  These two types of absorption components
are certainly produced by different mechanisms. We discuss them
separately.

\subsection{The narrow absorption components} 

The photon flux from hot stars drops sharply above 54 eV because of
the \heii\ ionisation edge, so photoionisation by hot stars can 
produce \siiv\ (33.5 eV) and \civ\ (47.9 eV) but not
significant \nv\ (77.5 eV) or \ovi\ (113.9 eV). This energetic
argument was, for example, advanced by \citet{fox03} to explain the
narrow \civ\ and \siiv\ absorption components observed towards a
Galactic sight line.  We show in the previous section that the
broadenings of the \civ\ and \siiv\ imply temperatures of a few $10^4$
K and likely less in the stronger features because of unresolved
components.  The velocities of the narrow  \civ\ and \siiv\ components 
are consistent in two cases with the velocities of
the \hii\ region or the expanding shell as traced by H$\alpha$ emission and
\feiii\ absorption. These observed properties are in agreement with
photoionisation playing an important
role in the production of these ions along with \feiii\ and H$\alpha$.
But in the two other cases, the bulk of the absorption in
\civ\ and \siiv\ do not correspond to either the systemic
or expanding velocities of the regions probed, suggesting  that additional 
or different production mechanisms are in play.

The \civ/\siiv\ ratios in the narrow components appear always greater than 1 (from AOD and
profile fitting comparisons), although we remind the reader that the
\siiv\ continua in the spectra of Sk$-$67\degr211 and
Sk$-$71\degr45 are uncertain (see \S\ref{civmeas}).  For Sk$-$67\degr211, 
Sk$-$69\degr246, and Sk$-$71\degr45, the hidden complexity
within the strongly-saturated \civ\ and \siiv\ profiles are likely
to have more than one component, and our column densities are thus
uncertain.  While the \ovi/\civ\ and \ovi/\siiv\ ratios are uncertain,
they appear to always be smaller than unity, implying very little
\ovi\ is produced by the same mechanisms that produce the strong,
narrow components of \civ\ and \siiv. This is further strengthened 
by the high \civ/\nv\ ratios (see footnote~\ref{foot-nv}).

Photoionisation models of diffuse \hii\ regions generally produce
ratios \civ/\siiv\ less than unity, although the addition of X-rays
can boost this ratio \citep[][]{black80,cowie81,knauth03}.  The
$b$-values of the narrow \civ\ and \siiv\ components seen in our data
strongly suggest, however, that photoionisation may play a role in
producing this material.  We have undertaken simple photoionisation
models using the Cloudy code \citep[v06.02.09; last described
by][]{ferland98} to test the assumption that the underlying structures
may give rise to the narrow components.  We calculated a model
spherical \hii\ region about a single star having properties similar
to the star Sk$-$65\degr22: $\log L/L_\odot = 6.2$, $T_{\rm eff} \approx
40000$ K \citep[from][]{massa03}.  We adopted LMC metal abundances
from \cite{russell92}.  Our models include a treatment of interstellar
grains for heating/cooling and opacity purposes; this dust
affects the abundances due to elemental depletion (see
Cloudy documentation).  We assumed a clearing within the inner-most
parsec of the \hii\ region and constant density outward of this radius.  
In general we find that such models, with an extremely bright, high
temperature star, can reproduce (or exceed) the measured \civ/\siiv\
ratios observed in the narrow components along the sight lines in this
work.  We examined models without the depletion (i.e., a non-self
consistent model with grains for thermal and opacity effects, but leaving all
the metals in the gas phase), which verified the depletion effects did
not dominate the \civ/\siiv\ results.  These single star \hii\ region
models, while able to match or exceed our observed \civ/\siiv\
measurements, do not simultaneously match this ratio and the total
column densities of the highly ionised species seen towards Sk$-$65\degr22.  In fact, the
models that produce a sufficiently high column density tend to
over-predict the \civ/\siiv\ ratio.  However, these are extremely
simple models that do not take into account the other stars in the
vicinity of our target objects, the complex geometries and density
structures within the regions, etc.  Given the high temperatures and
the quite high luminosities of our target stars, as well as their
locations within star forming regions in the LMC, it seems likely that
direct photoionisation from OB-type star photospheres could explain the
narrow \civ\ and \siiv\ along some sight lines.

Given the kinematic complexity of the other sight lines (see \S\ref{sec-kin}), 
several processes may be at work or that the structures giving rise to the
absorption have complex internal kinematics. In particular, early-type 
stars with strong stellar wind create large cavity in the ambient ISM 
\citep{castor75}. Bubbles have complex kinematics and have 
a mixture of hot ($T>10^6$ K) and cool ($T< 10^4$ K) \citep{weaver77}. 
The hot gas is principally found in the shocked wind. The cooler gas 
mainly is found in the shell
behind the conducting front following a rapid radiative cooling 
\citep[see also,][]{shapiro76a,sutherland93}. Such process also 
creates an interface between the hot and cool gas at $T\sim 10^5$--$10^6$ K where 
some of the observed \ovi\ is produced (see below). The ionisation of the outer shell depends 
mostly on the radiation of the central star \citep{weaver77}. It is interesting to note
that the two hotter stars in our sample show the strongest \civ\ and \siiv\ 
absorption associated with an expanding shell (see \S\ref{sec-kin}). The stellar wind
model therefore provides a qualitative explanation for part of 
observed narrow component associated with the outer/expanding shell. This is also 
supported by the recent {\em Chandra}\ observations of 30 Do, which 
shows that Sk$-$69\degr246 may be a colliding-wind binary with bright X-ray emission from 
2.1 keV plasma \citep{townsley06}. However, 30 Dor and other regions of 
the LMC are also well known for their superbubbles created not only 
by stellar winds but also SNRs. The SNR models of \citet{slavin92} can 
also produce cool highly ionised at $10^4$ K. This is not surprising since 
the SNR and stellar wind models explore very similar physics; in the former the source of 
the shocked is the supernova, in the latter case, it is the central star. 
The stellar wind and SNR models place the expanding shell within a radius of about 
30 to 100 pc from the central region. 
Yet, other physical scenarios might produce such cool highly-ionised gas,
as well.  For example, emission of a hot, cooling plasma could photoionise the gas; 
\ovi\ and \nv\ are negligible with respect to \civ\ and the temperature of the gas is 
low ($T<10^4$ K) so the \civ\ and \siiv\ broadenings are small \citep{knauth03}.

In summary, in view of the kinematical differences between the various 
sightlines, the origins of the narrow components observed in the  
\civ\ and \siiv\ profiles are most likely multiple. In the spectra of Sk$-$67\degr211
and Sk$-$71\degr75, it is likely that the narrow components of \civ\ and \siiv\ are in 
fact at least two narrow components (hence tracing gas at $T\la 10^4$ K) at the systemic 
and expanding shell velocities. In the spectrum of Sk$-$65\degr22, the
absorption is comparatively weak and consistent with a single narrow component
associated with the \hii\ region around the star. And in the spectrum of 
Sk$-$69\degr246, there is only evidence of a strong absorption at the 
velocity of the expanding shell, showing the peculiarity of this line of sight. 
While the exact origins of the highly ionized species probed by the
narrow components remain unknown, their properties can be chiefly 
associated with the stellar environments. 

\subsection{The broad absorption components}
Towards each line of sight, the \ovi\ LMC component is very broad with
$b \sim 50$ \km, much larger than the broadening of the Galactic plus
the IVC or the temperature at which \ovi\ peaks in abundance in CIE,
implying that the LMC \ovi\ component is composed of several kinematic
components. There is evidence for broad \civ\ and \siiv\ LMC
absorption components associated with the \ovi\ towards at least 3
lines of sight.  Although the velocity structure of the broad absorption 
is not fully known, the high-ion profiles suggest that this gas traces hot
collisionally ionised gas at temperatures of a few times $10^5$ K.  We also
showed in \S\ref{sec-kin} that \ovi\ and \siii\ have absorption over nearly
the entire range of velocities from $-50$ to $+330$ \km.  \civ\ and
\siiv\ cover a smaller velocity range.  In particular they do not
show absorption at the lowest or highest velocities associated with
the LMC.  Thus, gas at the lowest and highest velocities makes a
proportionately smaller contribution to the total columns of \civ\ and
\siiv\ than it does in \ovi.

\subsubsection{Ratios of the highly ionised species}

\begin{figure}
\includegraphics[width = 8.5truecm]{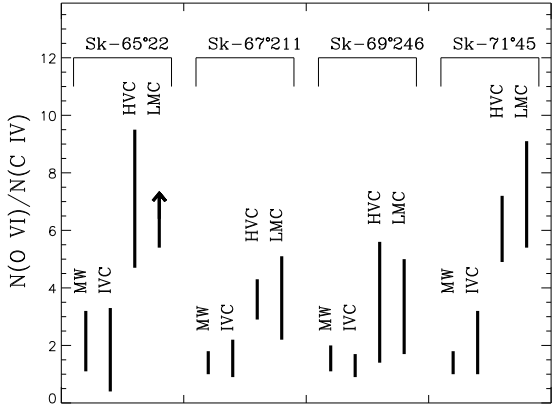}
\caption{Summary of the column density ratios of \ovi\ to \civ\ for the 
MW, IVC, HVC and broad LMC components towards the four sight lines, obtained from
the AOD measurements (see Table~\ref{t5}). \label{fig8} }
\end{figure}

Fig.~\ref{fig8} shows the \ovi/\civ\ ratio (from Table~\ref{t5}) along
our four sight lines, separated into the four principal components:
the MW ($v_{\rm LSR} \la +30$ \km), IVC ($30 \la v_{\rm
  LSR} \la 100$ \km), HVC ($100 \la v_{\rm LSR} \la
175$ \km), and broad LMC ($v_{\rm LSR} \ga 175 $ \km) velocity
components. 
Over the velocities where the narrow components discussed above are present the
ratios are inferred mostly from the profile fitting (Table~\ref{t4}),
except towards Sk$-$67\degr211 and Sk$-$69\degr246 where the absorption
is weak enough that the ratios of the apparent column densities can be
estimated over a portion of the absorption profiles.
In the broader wings of the absorption profiles, the ratios are
inferred from our AOD analysis (Table~\ref{t5}). For Sk$-$65\degr22,
the fits to the broad components of \civ\ and \siiv\ absorption appear
secure, but the fits to the broad absorption along the other
sight lines are not unique (see \S\ref{civmeas}). We found a good
agreement between the ratios estimated from the AOD and profile
fitting for Sk$-$67\degr211 and Sk$-$69\degr246, strengthening
the results from the profile fitting.

While there is some room for variations in the \ovi/\civ\ ratio as 
a function of velocity along a given sight line (our approach to deriving the values
plotted in Fig. \ref{fig8} is relatively conservative), the HVC and broad
LMC components have quite similar \ovi/\civ\ ratios, and the MW and IVC
components are generally in agreement with one another, but smaller than the HVC--LMC
values.  The observed \ovi/\siiv\ ratios for the LMC and HVC are
roughly consistent with those found in the MW or IVC components, but the \ovi/\civ\
ratios are generally quite a bit larger in the HVC and LMC material
than in the MW or IVC material.  Moreover, the \ovi/\civ\ ratios in HVC
and LMC gas seem to vary between sight lines, but the ratios in the
HVC and LMC components {\em along a given sight line} are always
very similar.

\subsubsection{Evidence for outflows}

The similarity of the \ovi/\civ\ ratios between the HVC and LMC
gas suggests that (i) the HVC component has its origins in the
LMC, and (ii) the processes producing the highly-ionised gas in the
LMC affect a large range of velocities along a given sight line in a
similar manner.  The first of these conclusions supports a conclusion
drawn by \citet{staveleysmith03} in their study of the \hi\ structure of
the LMC.  These authors found that the relatively high column density
clouds at velocities $v_{\rm LSR} \sim \,$100--160 \km\ towards the LMC are
often seen projected onto \hi\ voids in the LMC disk, and they are
connected to the disk with spatial and kinematic bridges in
position-velocity plots.  They also note the presence of material at
velocities outside the range of  velocities of the LMC disk.  They conclude
based upon these lines of evidence that the gas in these clouds is
likely expelled from the LMC disk.

Our absorption line measurements are much more sensitive to low column
density matter than the \hi\ emission line measurements of
\citet{staveleysmith03}.  Thus, we are potentially probing gas with
significantly different ionisation states.  Indeed, Staveley-Smith et
al. only display maps for the HVC material with $N(\mbox{\hi}) \ga
10^{18}$ cm$^{-2}$, which shows most of the gas concentrated to the
southwestern edge of the LMC.  Our lines of sight are concentrated
significantly further to the east than most of their clouds.
\citet{danforth02}, however, have demonstrated that HVC material can be
found over a substantial fraction of the LMC using {\em FUSE}\ absorption
line measurements.  Their measurements of \feii\ equivalent widths towards 
LMC stars show a broad swath of
high-velocity material covering the eastern edge of the galaxy, with
the highest equivalent widths arising in the northeast
quadrant.  (Note that the coverage of the southwestern quadrant in
their data was minimal.)

The most promising way to confirm our conclusion that the HVC material
is associated with the LMC is to obtain absolute abundances of the HVC
gas, which will require more sensitive \hi\ emission line
observations towards these sight lines.  An absolute abundance derived
from the ratio of \oi\ to \hi\ similar to the abundance of the LMC
would strongly suggest an LMC origin.  

While we acknowledge the uncertainties in this conclusion, we
nonetheless proceed assuming that the HVC towards the LMC is
associated with this galaxy.  Thus, the velocity range for LMC
gas along these sight lines should be defined as everything at
$v_{\rm LSR} \ga 100$ \km.  The velocities of the LMC IVC and HVC
components indicate that they are expanding away from the warm disk
gas or, for the gas at lower absolute velocities relative to the LMC
disk, that the gas is extended above the disk, perhaps rotating in a
thick disk. Only towards Sk$-$71\degr45 is there evidence for both 
significantly red-shifted and blue-shifted components in the weakly ionized
species and \ovi.  Thus, little of the material we see can
represent gas returning to the LMC disk.  The velocity of the HVC
relative to the LMC disk is larger than a simple estimate of the
escape velocity of the LMC, $v_{esc} \approx \sqrt{2} v_{circ} \sim
100$ \km\ \citep[with $v_{circ} = 72\pm7$ \km\ from][]{alves00},
implying that the material may escape the LMC, polluting the
intergalactic space between the LMC and the Milky Way or serving as
fuel for the Magellanic Stream.

The LMC-IVC may trace an interstellar thick disk about the LMC,
perhaps similar to that seen about the Milky Way 
\citep[e.g.,][]{savage97}.  However, it seems unlikely that the HVC material can be
associated with an extended, dynamically-supported thick disk
component due to its extremely large velocity separation from the LMC.
Theoretical modeling and stellar observations of the LMC have revealed
the existence of a stellar thick disk
\citep[e.g.,][]{weinberg00,bekki05,cole05}.  Neither the simulations
nor the observations of evolved stars demonstrate velocity dispersion
greater than about 20--40 \km.  While thick disk gas may not follow
the stellar kinematics, it is unlikely that the velocity dispersion of
the gas would be larger than that of stars since the gas is more
subject to hydrodynamical effects than stars (i.e. gas is not
collisionless).  Gas corotating in a thickened disk may appear at
lower velocities due to projection effects associated with the
inclination of the LMC disk to our line of sight
\citep[see][]{howk02}.  While this could contribute to the velocity
shift of the LMC-IVC relative to the thin disk material, it does not
produce a large enough velocity shift to explain the HVC velocities.
Thus, the HVC material seems unlikely to be associated with a thick
disk of material about the LMC.  It is more likely tracing a
large-scale outflow of material.

\subsubsection{Physical origins}

The comparison of the column densities of the highly ionised species can also 
be used  to unravel their origins. The models summarized in Table~\ref{t5} can 
be divided in 3 main mechanisms \citep{spitzer96}: (i) conductive heating mechanisms
(including the SNR, stellar winds, conductive interface, and shock ionisation) in 
which cool gas evaporates into adjacent hot medium (the highly ionised gas at 
$T\sim {\rm a\, few}\times 10^5$ being found at the interface between the cooler and hotter 
plasmas); (ii) radiative cooling models in which hot gas cools, (iii) the turbulent
mixing layers where hot gas mixes with cool gas. The broad component of the highly
ionised species therefore always probes indirectly the presence of hot gas ($T>10^6$ K).
In real physical environments, it is likely that these mechanisms are actually mixed all 
together to a certain  degree, but depending on the physical conditions, one may dominate 
the other. 

We discussed in the previous section that strong stellar winds (or SNRs) most certainly 
occur,  at least in the bright \hii\ and superbubble regions. Therefore some 
of the \ovi\ and broad \civ\ and \siiv\ absorption must originate in the interface
between the shocked plasma and the swept-up interstellar gas. 
In velocity space, most of this absorption should occur near the expanding 
shell velocity where part of the strong narrow components \civ\ and \siiv\ 
are observed. While the observed high-ion ratios are very roughly consistent with 
those models, the observations also suggest that the environments very local to our
target stars have little discernible effect on the
\ovi/\civ\ ratios of the broad absorbing components.  The \ovi/\civ\ ratios are the highest
towards the faint \hii\ region associated with the supergiant shell
DEM\,48 (Sk$-$65\degr22) and towards the superbubble DEM\,221
(Sk$-$71\degr45). They are the lowest towards the \hii\ region DEM\,241
(Sk$-$67\degr211) and the X-ray bright superbubble 30\,Dor
(Sk$-$69\degr246). This difference between the two superbubble sight
lines and the two \hii\ region sight lines makes it impossible to
conclude that one type of structure is responsible for producing
higher or lower \ovi/\civ\ ratios. The absence of
clear signature in the high-ion ratios may be because
the individual clouds cannot be distinguished. 
A very large fraction of the observed \ovi\ (and broad component
of \civ\ and \siiv) must also have a different origin than the stellar environment 
itself (stellar wind and SNR models predict $N($\ovi$)\sim {\rm a\, few}\, \times 10^{13} $ cm$^{-2}$, 
much less than the observed total \ovi\ column density)
and is not spatially localized near these \hii\ regions and superbubbles
even though some of it overlaps these structures in velocity space. 

When the observed LMC-HVC high-ion ratios are compared to theoretical models 
(see Table~\ref{t5}), the turbulent mixing layer (TML) model
always fails to produce large enough \ovi/\civ; this is at odds with
gas at the Milky Way velocities for which TMLs seem to be consistent
with the observed ratios of highly ionised species (see Tables~\ref{t4} and \ref{t5}). 
One caveat to this conclusion is that if the oxygen abundance is systematically enhanced with 
respect to carbon, it would be likely that more \ovi\ is observed than
in ``typical" regions in the Milky Way. But  enhancement of oxygen
or deficiency of carbon are not observed in the LMC (see Appendix in Welty
et al. 1999, and references therein).

\begin{table*} 
\begin{minipage}{11 truecm}
\caption{\large Interstellar \ovi\ and environments in the LMC \label{t7}}
\begin{tabular}{clcccl}
\hline
Id.    &   Star & R.A.& Dec. & $\log N($\ovi) &  Environments \\
&  &  (\degr)&  (\degr)&  (dex)&    \\ 
\hline
1  & Sk$-$67\degr05	&  72.58  &	--67.66  &   $  13.89 \pm 0.07	  $	  & \hii\ region DEM\,7 \\
2  & Sk$-$67\degr20	&  73.88  &	--67.50  &   $  14.26 \pm 0.10	  $	  & Field star, no H$\alpha$ \\
3  & Sk$-$65\degr22	&  75.34  &	--65.87  &   $  14.14 \pm 0.02	  $	  & Supergiant shell DEM\,48 \\
4  & Sk$-$66\degr51	&  75.79  &	--66.68  &   $  14.31 \pm 0.07	  $	  & \hii\ region DEM\,56, no H$\alpha$ \\
5  & Sk$-$67\degr69	&  78.58  &	--67.13  &   $  14.48 \pm 0.03	  $	  & \hii\ region DEM\,107 \\
6  & Sk$-$68\degr80	&  81.62  &	--68.84  &   $  14.61 \pm 0.04	  $	  & X-ray bright superbubble DEM\,199 \\
7  & Sk$-$70\degr91	&  81.89  &	--70.61  &   $  14.55 \pm 0.07	  $	  & Superbubble DEM\,208 \\
8  & Sk$-$66\degr100   	&  81.94  &	--66.92  &   $  14.26 \pm 0.06	  $	  & Field star, no H$\alpha$ \\
9  & Sk$-$67\degr144   	&  82.55  &	--67.43  &   $  14.41 \pm 0.07	  $	  & Periphery of supergiant shell LMC\,4 \\
10 &  Sk$-$71\degr45   	&  82.81  &	--71.06  &   $  14.37 \pm 0.02	  $	  & Superbubble DEM\,221 \\
11 &  Sk$-$69\degr191  	&  83.58  &	--69.75  &   $  14.38 \pm 0.06	  $	  & X-ray bright superbubble DEM\,246\\
12 &  Sk$-$67\degr211  	&  83.80  &	--67.55  &   $  14.20 \pm 0.02	  $	  & \hii\ region DEM\,241 \\
13 &  BI\,237	 	&  84.06  &	--67.65  &   $  14.39 \pm 0.03	  $	  & Diffuse \hii\ region \\
14 &  Sk$-$66\degr172  	&  84.27  &	--66.35  &   $  14.31 \pm 0.06	  $	  & \hii\ region DEM\,252\\
15 &  Sk$-$69\degr246  	&  84.72  &	--69.03  &   $  14.36 \pm 0.08	  $	  & X-ray bright superbubble 30\,Dor \\
16 &  $<$N\,70$>$  	&  85.81  &	--67.85  &   $  14.48 \pm 0.10	  $	  & Superbubble DEM\,301 \\
17 &  CAL\,83	 	&  85.89  &	--68.37  &   $  14.39 \pm 0.08	  $	  & Field star \\
18 &  BI\,272	 	&  86.09  &	--67.24  &   $  14.30 \pm 0.03	  $	  & Diffuse \hii\ region DEM\,309 \\
19 &  Sk$-$67\degr 266  	&  86.46  &	--67.24  &   $  14.09 \pm 0.07	  $	  & Shell  DEM\,315 \\
\hline
\end{tabular}
Note: \ovi\ column densities are from this work, \citet{howk02}, and \citet{danforth06}. 
The designations DEM refer to entry in the catalog of \citet{davies76}.
\end{minipage}
\end{table*}

\begin{figure*}
\includegraphics[width = 15truecm]{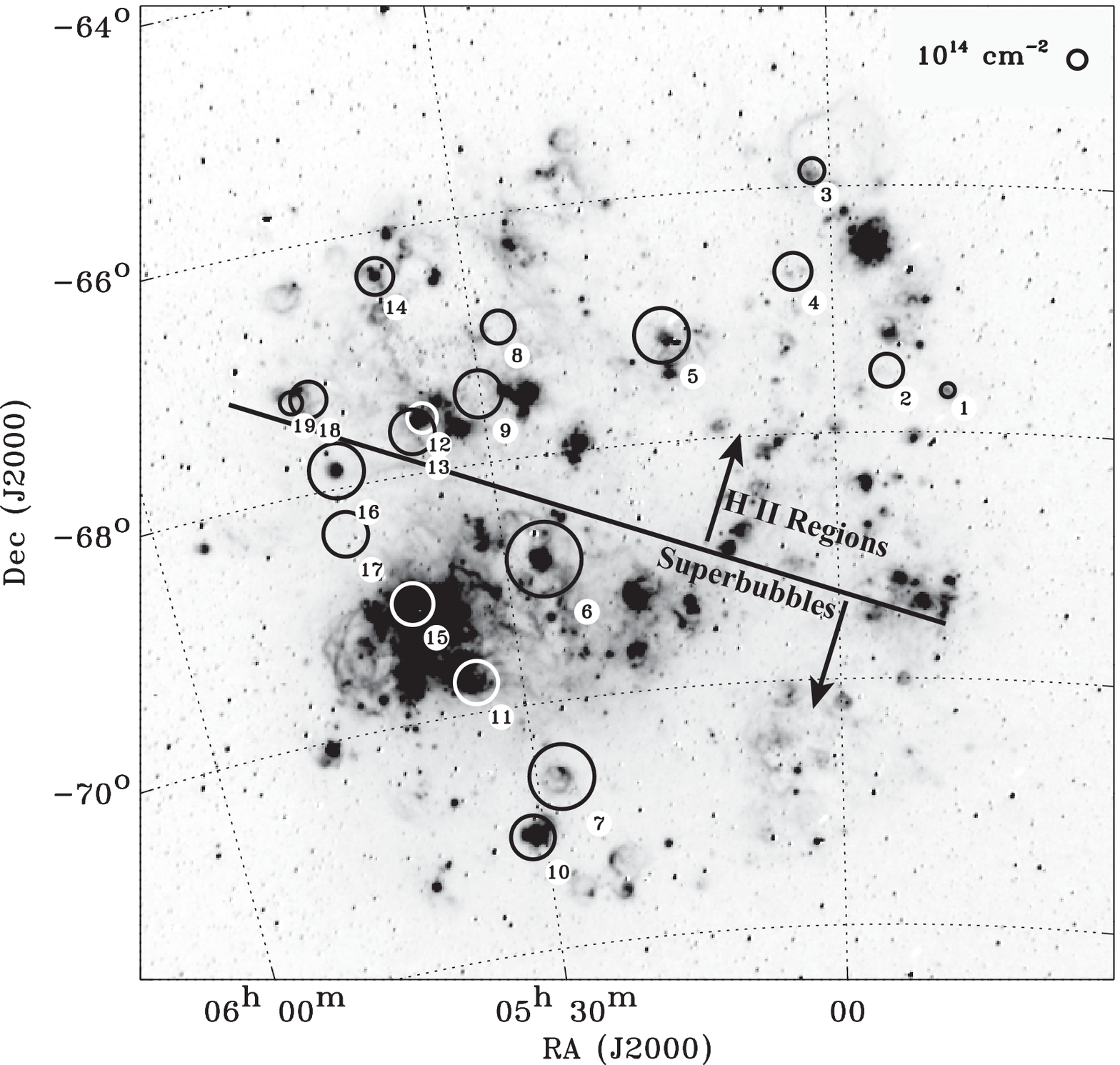}
\caption{H$\alpha$ image of the LMC \citep{gaustad01}. The
  identification numbers are summarised in Table~\ref{t7}. Darker regions correspond
  to brighter H$\alpha$ emission. The radius of the circle marking
  each probe star is linearly proportional to the column density of
  interstellar \ovi\ at LMC velocities. {\em Note that the apparent
    emission near no. 5 is caused by an imperfectly subtracted
    foreground star}. Fortuitously, the superbubbles are well 
    separated from the \hii\ regions (except for no. 17, which is an \hii\ region)
    and the solid line marks the separation between them. Note that 
    several \hii\ regions are associated with supergiant shells. 
     \label{fig9}}
\end{figure*}

Another systematic between the various sightlines is deduced from Figs.~\ref{specmodel} 
and \ref{specmodel1} (see also discussion in \S\ref{sec-kin}): 
cooler gas traced by \siii\ and \oi\ is observed at all velocities
where \ovi\ absorption is detected.  The close correspondence in the
total velocity extent of the observed \siii\ and \ovi\ absorption strongly
suggests that the high and the weakly ionised species are kinematically coupled.  A
correspondence between \ovi\ and low-ion velocity extent has also been noted
in the Milky Way \citep{howk03,bowen06}.  Such a relationship is
expected if much or all of the \ovi\ arises in interfaces between warm
and hot gas. \citet{borkowski90} present models of conductive
interfaces in such an arrangement; they predict that if the \ovi\ is found in both the
evaporative and conductive stage, the \ovi\ column density would be
less than a few $10^{13}$ cm$^{-2}$, implying that each line of sight
in our work intercepts 5--10 interfaces. The broadening of the
high-ions is consistent with such an interpretation since, if there
are 5--10 interfaces, each component of \ovi\ could still have a
temperature of a few times $10^5$ K. While the low
ion profiles are mostly unresolved, their velocity extents are consistent
with a large number of components, and therefore with large number
of interfaces.  The profiles of weakly and highly ionised species cover a large enough 
velocity range that they are likely not simply tracing components
associated directly with the \hii\ regions or superbubbles housing the
target stars.  Thus, multiple interfaces are certainly possible and
likely associated with gas out of the disk of the LMC.  If conductive
interfaces are the main origin for the broad component of the high
ions, they have to be at least $3\times 10^5$ year old to reproduce
the high-ion ratios. We note that the observed \ovi/\siiv\ ratios are
somewhat smaller than the ratio predicted by \citet{borkowski90}, but
\siiv\ can be easily photoionised by other sources (e.g., photons from
stars or the absence of the incorporation of \heii\ $\lambda$304
photons in the interface model that are capable of ionising efficiently
\siiii\, see Savage, Sembach, \& Cardelli 1994), and this is not taken into account in 
their models. The limits derived from \nv\ are always consistent with the conductive interface
models.  It is not clear within the frame work of the conductive interface models
why the high-ion ratios would be similar over the large velocity range
between the LMC and HVC components, and yet so dissimilar between the
observed sightlines.

In view of the complexity of the profiles and of the physical 
regions probed, other models may be in play, especially in the
event that the apparent kinematic coupling does not imply
coincidence of the highly and weakly ionised gas:

-- Strong shocks with $v_{\rm shock} >200$ \km\ would produce enough
\ovi\ to explain the observed ratios. Such strong shocks have not
been inferred in the denser parts of \hii\ and superbubble regions of the LMC
\citep{oey00}. But high-velocity shocks are likely to arise in low-density
gas where they remain undected since they would not contribute to the 
nebular emission (Y.-H. Chu 2006, priv. comm.).
The strong shock models are attractive because 
they provide a straightforward method for producing a consistent 
\ovi/\civ\ ratio over a large range of velocities. 
If such strong shocks occur in the LMC, the gas traced
by the LMC-IVC/HVC components could have been ejected away from the LMC in
this process.

-- The non-equilibrium radiative cooling model of \citet{edgar86} where a gas cools from
temperature of $\ga 10^6$ K fits also the data when \ovi/\civ\ is
greater than 5.9.  More recently, \citet{gnat06} have produced new
calculations of radiatively cooling gas as a function of metallicity
using updated atomic data.  The \ovi/\civ\ and \ovi/\siiv\ ratios for
the HVC and LMC absorbers are consistent with the Gnat \& Sternberg
models of radiative cooling from $T\sim5\times10^6$ K for models with
metallicities 0.1 to 1.0 times solar.  The gas must still be at
temperatures between $\sim(1-2)\times10^5$ K in order to match their
models.  A radiative cooling model for the origins of the
highly-ionised gas in the HVC and LMC components would be consistent
with an outflow and infall of highly ionised material from the disk of
the LMC, i.e., an exchange of matter between the thin disk of the LMC 
and its halo or thick disk. The majority of the gas must cool in the
outflow phase since we only find evidence for the possible infall of
material towards one sight line.  If we combine our sample with
that of \citet{howk02}, only 3 of 14 sight lines show possible
evidence for infalling material. It is, however, not clear within the
framework of this model why velocities of the cool and hotter gas
would be systematically coincident.

In summary, the physical origins of \ovi\ and broad \civ\ and \siiv\ 
are tangled and several processes could occur. However, it is apparent that 
conducting heating processes play a fundamental role in the production 
of the highly ionised species probing gas at $T\sim 10^5$ K
while TMLs are not important in the probed regions in the  HVC-LMC,
except if a large fraction of \ovi\ is produced via another mechanism (e.g., 
radiative cooling).

\subsubsection{Fluctuation of $N($\ovi)}

We note that \citet{danforth06} found that the \ovi\ column densities
appear higher by about 0.2 dex in the N\,70 superbubble and other
superbubbles described by \citet{howk02} compared to some more
quiescent \hii\ region sight lines, suggesting the very local
environment of the background stars may affect the total column of
\ovi\ along a sight line.  In this context, they argued that
superbubbles were significant contributors to the total \ovi\ observed
along LMC sight lines.  Fig.~8 shows an H$\alpha$ image of the LMC
from \citet{gaustad01} with a representation of all available \ovi\
column densities superimposed \citep[including][and this work]{danforth06,howk02}.  
Table~\ref{t7} summarises the measured
columns and the surrounding environments for each {\em FUSE}\
measurement.  All of the superbubble sight lines are concentrated
below the line superimposed on the image (we display only the average
of the four measurements towards N70).

While some superbubble-type regions may have slight enhancements in
\ovi, what is striking about this figure is the very large local
variations, as reported by \citet{howk02}. For example the pairs of
stars labeled 12 \& 13 (Sk$-$67\degr211 \& BI~237) and 18 \& 19 (BI~272
and Sk$-$67\degr266) trace \hii\ regions where the column densities
varies by a factor 2 from 14.1 to 14.4 dex.  These pairs have
projected separations of 90 and 129 pc, respectively.  This is the
degree of enhancement \citet{danforth06} identify with superbubbles,
but over much smaller angular scales than those authors use for
comparison between superbubble and non-superbubble sight lines.
Furthermore quiescent regions like stars no. 2, 5, and 17 have \ovi\
column densities similar or larger than observed in some superbubbles.
The dispersion of $N(\mbox{\ovi})$ is large along superbubble sight
lines, as well.  The measurements of $N($\ovi) appear not to be
smaller than 14.3 dex in the superbubbles.  However, many more \ovi\
measurements will be needed in the LMC before one can truly determine
if the superbubbles and other very local phenomena have a strong
impact on the observed total \ovi\ column density.  It is clear that the
\ovi\ has a patchy distribution in the LMC, and comparisons between
sight lines separated by tens of parsecs may not lead to significant
results in this regard.  While the high-ion ratios vary from sight
line to sight line, there is no clear link between the observed
variation and the regions associated with the lines of sight. The
comparison of high-ion ratios is potentially a more secure indicator 
of the impact of the local effects because it is
not subject to the varying depth of the gas probed by the background
stars, which may be part of the source of the observed variations in
$N($\ovi).

\section{Summary}\label{sec-sum}
We analyse the physical properties of the highly ionised species (\civ, \siiv, \nv, \ovi)
in and around the LMC towards a sample of 4 LMC hot stars. Two of
these sight lines probe prominent superbubbles, while the other two
probe \hii\ regions, allowing us to compare the highly ionised species and their 
connection with weakly ionised species among different regions. \ovi\ was 
obtained with  {\em FUSE} with a spectral resolution of $\sim$20 \km, and 
the other highly ionised species were observed with the {\em HST}/STIS E140M setup with a spectral resolution 
of $\sim$7 \km. The excellent spectral resolution and high signal-to-noise of these data
have allowed us to model the highly ionised species in the LMC. 
Our main findings are summarised as follows:

(i) Absorption from both the high and low ionisation species are 
observed over the entire range of velocities from the Milky Way to LMC ($-50 \la v_{\rm LSR} \la +350$ \km).
Along our four sight lines, the profiles are separated into four principal components:
the MW ($v_{\rm LSR} \la 30$ \km), IVC ($30 \la v_{\rm
  LSR} \la 100$ \km), HVC ($100 \la v_{\rm LSR} \la
175 $ \km), and LMC ($v_{\rm LSR} \ga 175 $ \km). \nv\ is not detected 
at the 3$\sigma$ level. 

(ii) In the LMC component, we find narrow and broad absorption 
in \civ\ and \siiv, but only very broad \ovi\ absorption. 
The narrow LMC components of \civ\ and \siiv\ 
have an observed $b$-value that implies temperatures of a few times $10^4$ K or less. 
The properties of the narrow components of \civ\ and \siiv\ can be
principally explained by the interstellar environment associated with
the stars, in particular \hii\ regions around the central star 
and expanding shells behind the shocked plasma. 

(iii) The \ovi\ at LMC velocities is
broader than the \ovi\ absorption observed in the combined
Milky Way and IVC component, implying that the LMC \ovi\ profiles are 
composed of several components. The broad LMC components of \ovi, \civ, and \siiv\ 
likely probe gas of a few times $10^5$ K.

(iv)  We find a striking similarity in the \ovi/\civ\ ratios 
for the LMC and HVC components ($v_{\rm LSR} \ga 100$ \km) along a  given sight line, but the  
\ovi/\civ\ ratios for the HVC and LMC gas vary between the sight lines (see Fig.~7). 
The fact that the ratios in the HVC and LMC components along a given sight line are always
very similar suggests that the HVC component has its origins in the
LMC, and  the processes producing the highly-ionised gas in the
LMC affect a large range of velocities along a given sight line.

(v) The difference in the high-ion ratios between the four sight lines 
implies  different processes or varying conditions may produce the highly ionised species 
in the LMC. There is no evidence, however, of a link between the observed 
variation and the regions associated  with the stars, although it is 
likely that some of the \ovi\ absorption occurs because of stellar winds
or other phenomena associated with the stellar environment. 
Conductive interface model can reproduce both the high-ion ratios observed 
in the broad component and the apparent kinematically coupling between
\ovi\ and the weakly ionised species such as  \siii\ and \oi. 

(vi)  The velocities of the LMC-IVC and HVC components indicate that they are 
expanding away from the warm disk gas. Only towards one line of sight is there evidence 
for both significant red-shifted and blue-shifted components in the low- and
high-ion spectra, implying little of the material we observe is
returning to the LMC disk. Our analysis  therefore provides compelling 
evidence for a hot LMC halo fed by energetic outflows from the LMC disk and 
even possibly with a galactic wind since the velocity of the HVC relative to the LMC
disk is actually large enough to escape the LMC.

\section*{acknowledgements}

We are grateful to You-Hua Chu and Blair Savage for useful comments 
on an earlier draft. We also thank You-Hua Chu
for sharing unpublished H$\alpha$ velocities.
We recognise partial support by the National Science Foundation grant 
AST 06-07731 and appreciate support from the University of Notre Dame. 
Based on observations made with the NASA-CNES-CSA 
Far Ultraviolet Spectroscopic Explorer. FUSE is operated for NASA by the Johns 
Hopkins University under NASA contract NAS5-32985. Based on observations made with the 
NASA/ESA Hubble Space Telescope, obtained at the Space Telescope Science Institute, which 
is operated by the Association of Universities for Research in Astronomy, Inc. under NASA
contract No. NAS5-26555. This research has made use of the NASA
Astrophysics Data System Abstract Service and the SIMBAD database,
operated at CDS, Strasbourg, France.

\bsp

\label{lastpage}

\end{document}